\documentclass[%
 reprint,
superscriptaddress,
 amsmath,amssymb,
 aps,
]{revtex4-2}

\usepackage{graphicx}
\usepackage{dcolumn}
\usepackage{bm}
\usepackage{xcolor}
\usepackage[normalem]{ulem}
\usepackage{tikz}
\usetikzlibrary{shapes.geometric, arrows}
\usepackage{nicematrix}

\begin{document}

\preprint{APS/123-QED}

\title{Design study of a low emittance complex bend achromat lattice}

\author{Minghao Song}
\email{msong1@bnl.gov}
\affiliation{Brookhaven National Laboratory, Upton, NY, USA, 11973}
\author{Timur Shaftan}
\affiliation{Brookhaven National Laboratory, Upton, NY, USA, 11973}

\begin{abstract}

Light sources worldwide have experienced rapid growth in the last decades, pushing towards higher brightness with lower emittance to meet growing demands from the user community. The quest for higher brightness motivates the development of low-emittance ring lattices. At this point, all fourth-generation storage ring light sources employ variations of the Multi-Bend Achromat (MBA) lattice. In this paper, we discuss an extension of this approach, known as Complex Bend\cite{Timur_2018_techreport} Achromat (CBA) lattice in relation to the future NSLS-II upgrade. A detailed approach for the lattice design will be described, and the developed lattice will be presented. The benefits of using a complex bend approach are demonstrated by achieving a small natural emittance of 23 pm at a beam energy of 3 GeV, straight sections of 8.4 m for long IDs acquiring a ratio of about 50\% of the drift space with respect to the ring circumference, compact ring elements (complex bends) based on Permanent Magnets and a large-scale reduction in the number of power supplies. Our new approach provides an extension to the MBA concept for the next-generation light source lattice design.

\end{abstract}

\maketitle

\section{Introduction}

Through the past few decades, synchrotron radiation light sources have become major hubs for modern applied science and have been driving research in various fields ranging from biology to physics, chemistry, and material science. As modern accelerator technology continues to expand its capabilities to produce bright radiation, the next-generation storage rings offer diffraction-limited X-ray beams for synchrotron light sources of tomorrow. 

The brightness of synchrotron radiation is one of the key characteristics of the source. It is defined by the electron beam emittance and intensity in combination with the chosen insertion device and the X-ray beam size and divergence in a source point. Following the quest to increase the source brightness, the ring designers seek ways to reduce the beam emittance.  
The electron beam emittance $\epsilon_{x0}$ defined by the storage ring lattice scales as $E^2/[N_cN_b]^3$, where $E$ is the beam energy, $N_c$ is the number of cells in a ring and $N_b$ is the number of bending magnets per cell~\cite{Lee_accel_phys_book}. Therefore, increasing the number of dipoles is the most effective way to reach ultra-low beam emittance.
This was recognized in the 90s by Einfeld et al.~\cite{Einfeld_DIFL}, and currently, every modern storage ring light source upgrade project is based on a certain variant of the multi-bend achromat (MBA) approach. This powerful concept has been detailed for MAX-IV~\cite{Leeman_MAXIV}, SIRIUS~\cite{Rodrigues_Sirius}, ESRF-EBS~\cite{Raimondi_ESRF-EBS}, APS-U~\cite{Fornek_APS-U}, ALS-U~\cite{Steier_ALS-U,Hellert_ALS-U}, HEPS~\cite{Jiao_HEPS}, DLSR~\cite{Raimondi_DLS} and customized at many other light source facilities~\cite{Streun_SLSII,Bartolini_Diamond-U,ElettraII}.

As one approaches the diffraction limit for the rings within a conceivable range of circumferences ($<$ 1~km), the challenge of available space for the ring elements becomes apparent and dominating. As an example of the ultimate lattice design for diffraction-limited source, a lattice option of the upgrade of MAX-IV~\cite{Leeman_MAXIV} features a 19-BA achromat, capable of reaching 16~pm rad in the 500-m long ring tunnel~\cite{Tavares_MAXIV_future}. This lattice provided a high-performance solution by reducing the beam emittance with a factor of 20 compared to the operating MAX-IV ring. However, the composition of the lattice elements left very limited space for insertion devices (IDs), vacuum components, and other equipment.

This example adequately illustrates the limit of the MBA approach in its current form. It motivates the search for new accelerator technologies and lattice solutions, which is the topic of this paper. 
The new accelerator methodology is presented by the concept of the complex bend element~\cite{Timur_2018_techreport}, a long-curved magnet containing a tight arrangement of quadrupole-dipoles built using permanent magnets (PM). The R\&D on PMQ magnets is an active area of research already delivering promising results at several facilities~\cite{Sushil_retreat_slides,Brooks_BNL_PMQ,N'gotta_ESRF_PMQ,Ghaith_SOLEIL_PMQ}. 

Our approach described in this paper is based on the concept of complex bend (CB)~\cite{Timur_Methodology_TCBA,Guimei_CBI,Guimei_CBII}. We aim to greatly increase the number of bends in the ring by superimposing dipole and quadrupole fields in the shortest CB segments, whose parameters are limited by the possible gradient in PMQs at the magnet bore size constrained by the size of the extraction channel for synchrotron radiation. This leads to a modification of the expression above $\epsilon_{x0} \propto E^2/[N_c N_{CB} N_b]^3$, where $N_c$ is the number of cells, $N_{CB}$ is the number of complex bends per cell and $N_b$ is the number of short dipole-quadrupole magnets per complex bend element.
To illustrate the advantage of this approach, we recall that the 7-BA APS-U lattice has a total of 280 dipoles in a ring (as compared with the APS lattice with 80 bends), while the complex bend lattice for NSLS-IIU could fit 30×3×30 = 900 dipoles. This opens an opportunity to reduce the emittance further than that of MBA and achieve a compact lattice arrangement that leaves ample space for machine elements and IDs.

\begin{figure}[htbp]
    \centering
    \includegraphics[width=0.4\columnwidth]{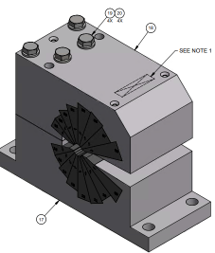}
    \includegraphics[width=0.4\columnwidth]{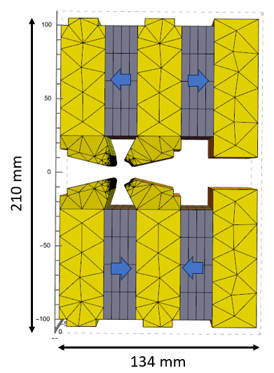}
    \caption{Left plot: Halbach PMQ; right plot: Hybrid PMQ.}
     \label{PMQ}
\end{figure}

A complete complex bend magnet is designed as an assembly of several combined dipole cells, with every cell providing a specific dipole field with alternating focusing and defocusing quadrupoles (QF-Drift-Bend-Drift-QD-Drift-Bend-Drift)~\cite{Guimei_CBI}. In~\cite{Timur_2018_techreport,Guimei_CBI}, we have shown that a much lower emittance could be reached if a regular DBA dipole is replaced by a complex bend element. Furthermore, we presented several versions of the complex bend magnet in~\cite{Guimei_CBI,Guimei_CBII}, where the bending was achieved by either introducing an external dipole field to the pure PM quadrupole field or shifting quadrupole magnets off-axis. In modeling such geometry, it has become apparent that this viable solution presents difficulties with achieving reasonable field quality at large transverse shifts of the magnets. The alternative solution that we recently developed and detailed is integrating quadrupole field into dipole field via a permanent magnet quadrupole (PMQ) of either Halbach or hybrid type\cite{Sushil_retreat_slides}, which as shown in Fig.~\ref{PMQ}. 
PMQ of either type will be employed in our design study of the complex bend lattice described below.

This paper complements ongoing efforts on designing the complex bend element and focuses on developing the lattice based on such elements. The optics that we describe here rely on a large number of short PMQs with the fields and strengths tailored to minimize Courant-Snyder invariant $\mathcal{H}$~\cite{Lee_accel_phys_book} while adjusting Twiss functions for their optimal values in the chromatic regions, matching sections and ID straights. 

Throughout the paper, we will demonstrate the benefits of the CB approach in optimizing the following five performance metrics in the lattice design:
\begin{enumerate}
 \item Low emittance;
 \item Long straights for IDs;
 \item Large space in the lattice available for accelerator equipment;
 \item Reduction of the number of power supplies and in power consumption;
 \item Optimal beam dynamics properties for the ring lattice leading to low-loss injection and long beam lifetime.
\end{enumerate}

In this paper, we will be considering complex bend lattice for NSLS-II upgrade (NSLS-IIU) as an example. 
NSLS-II is a third-generation light source with a circumference of 792 m. It contains 30 DBA cells and operates at 400 mA using top-off injection. The horizontal emittance is 2 nm for the bare lattice and can be reduced below 1 nm with three sets of damping wigglers. The future machine upgrade aims to increase brightness 10-fold at 1 keV to 50-fold at 10 keV of photon energy compared with the current facility.

\begin{figure}[htbp]
    \centering
    \includegraphics[width=\columnwidth]{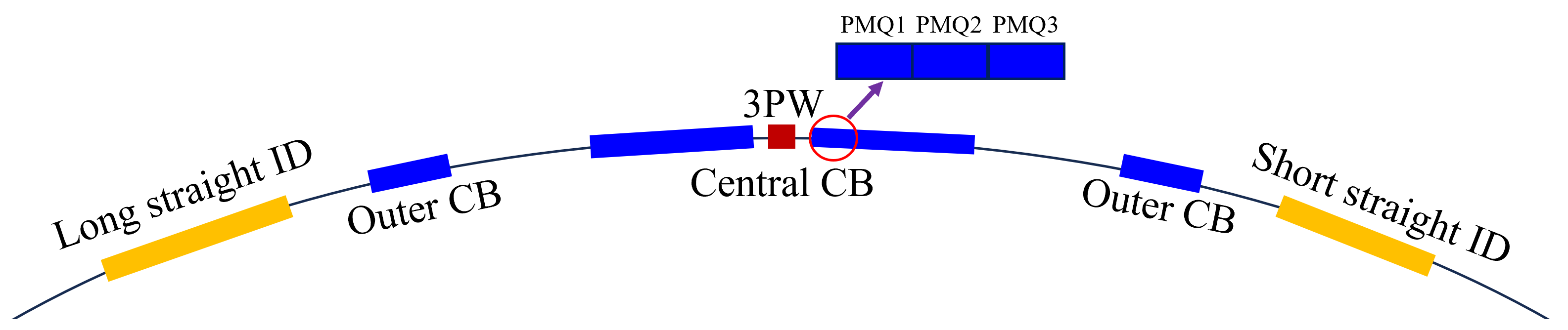}
    \caption{Magnet layout for a single cell of NSLS-IIU complex bend lattice.}
     \label{one_cell_mag_layout}
\end{figure}

Figure~\ref{one_cell_mag_layout} shows a single-cell magnet layout of the NSLS-IIU complex bend lattice candidate, which consists of one central CB and two outer CBs. The complex bends contain arrays of compact PMQ magnets. IDs in the long straight and short straight and 3-pole-wiggler (3PW) in the center of central CB provide the sources of the photon beam. 

The structure of this paper is organized as follows: Section~\ref{approach} describes the approach for the complex bend lattice design. Section~\ref{result} discusses the design of the complex bend achromat (CBA) lattice, one of the developed solutions, and shows the performance comparison between the CBA lattice and several selected MBA lattices. Section~\ref{conclusions} concludes the paper.

\section{The approach for lattice design}\label{approach}

The common goals of developing a light source lattice are to:
\begin{enumerate}
 \item Achieve a low-emittance solution;
 \item Demonstrate long straights available for IDs;
 \item Fit the lattice solution within the present facility tunnel satisfying tight constraints of overlapping ID and BM source points with these for the existing facility beamlines;
 \item Deliver a robust lattice design, resilient to errors and imperfections;
 \item Deliver a composition of the lattice elements with minimal power consumption;
 \item Rely on the maximal extent of feasible approaches in magnet, vacuum, and RF technology.
\end{enumerate}

In our approach, the search for the best lattice solution is converted into a “usual” accelerator methodology:
\begin{equation}
\label{OP_space}
\begin{aligned}
      O(\underbrace{\Delta O}_{\substack{\text{constraints:} \\ {J_x} \\{\alpha_c} \\ {\bm{L_s}} \\ {\bm{\beta_{ID}}} }})=\mathcal{M}(\underbrace{\Delta\mathcal{M}}_{\substack{\text{constraints:} \\ {geometry} \\{Twiss}}}) P(\underbrace{s, p}_{\substack{\text{variables:} \\ {strengths} \\{positions}}}, \underbrace{\Delta s, \Delta p}_{\substack{\text{constraints:} \\ {strengths} \\{positions}}}),\\
      \mathcal{F}=\mathcal{F}(O,weight) \text{ is the lattice figure of merit used for} \\ \text{selection of the best solution} 
\end{aligned}
\end{equation}
where $O$ are ``observables" or criteria for the lattice design (i.e., natural emittance $\epsilon_{x0}$, natural chromaticities $\bm{\xi}=(\xi_x, \xi_y)$, radiation loss per turn $U_0$, momentum compaction $\alpha_c$, lengths of long and short straight sections $\bm{L_s}=(L_{LS}, L_{SS})$, and beta functions $\bm{\beta_{ID}}=(\beta_{x,ID,LS}, \beta_{y,ID,LS}, \beta_{x,ID,SS}, \beta_{y,ID,SS})$ at the center of IDs). $\Delta O$ is the $O$-constraints set for the search of the best solution (i.e., ranges for $J_x$, $\alpha_c$, $\bm{L_s}$, and $\bm{\beta_{ID}}$). $P$-variables are the lattice parameters (i.e., strengths `s' and positions `p' of lattice elements). $P$-constraints are constraints for the strengths and positions, and $\mathcal{M}$ is a functional form between $O$ and $P$. In addition, there are constraints $\Delta\mathcal{M}$ related to the functional form $\mathcal{M}$. These contain geometrical limitations related to fitting to the ring tunnel and constraints imposed on the Twiss parameters. 

Below, we briefly describe our approach in the lattice design~\cite{Timur_Methodology_TCBA}. The whole lattice cell is divided into a number of bins. Every one of these bins is filled with one of several lattice elements, e.g., drifts, complex bends, and quadrupoles. The strengths of the CB elements are varied within the limits given by the gradient constraints defined in the $P$-space. The range of a single cell is split into several regions. Under the $P$ and $\mathcal{M}$ constraints, the $P$-variables of each region are optimized to achieve the best performance in $O$-space by using several optimization methods. The latter is based on a combination of MAD8~\cite{Iselin_MAD8}, Matlab~\cite{matlab}, and a suite of optimization algorithms.  
In the following, we describe in detail the principles of setting up the space of $P$, the space of $O$, and the methods to find the optimal solution. 

The solvability of the problem depends on the simplification of the functional form $\mathcal{M}$. It is known that optimization problems fail when the number of $P$-variables greatly exceeds the number of $O$-variables, which is the specific challenge for the problem that we studied. Below, we describe a method that improves conditions of the $P$-space where we use polynomial forms to reduce the dimension of $P$-vector.  

\subsection{Setup of the P-space}

\begin{figure}[htbp]
    \centering
    \includegraphics[width=\columnwidth]{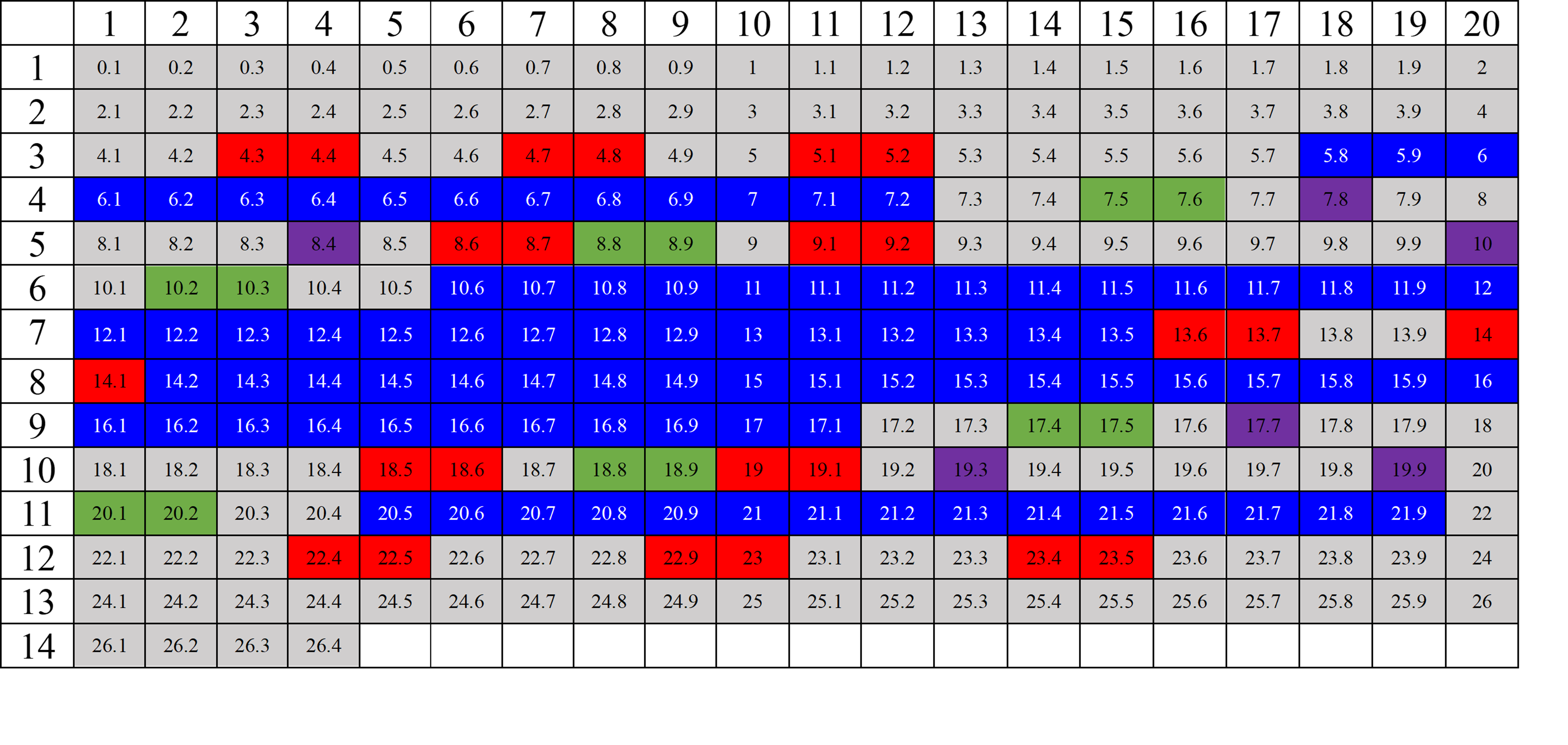}
        \includegraphics[width=\columnwidth]{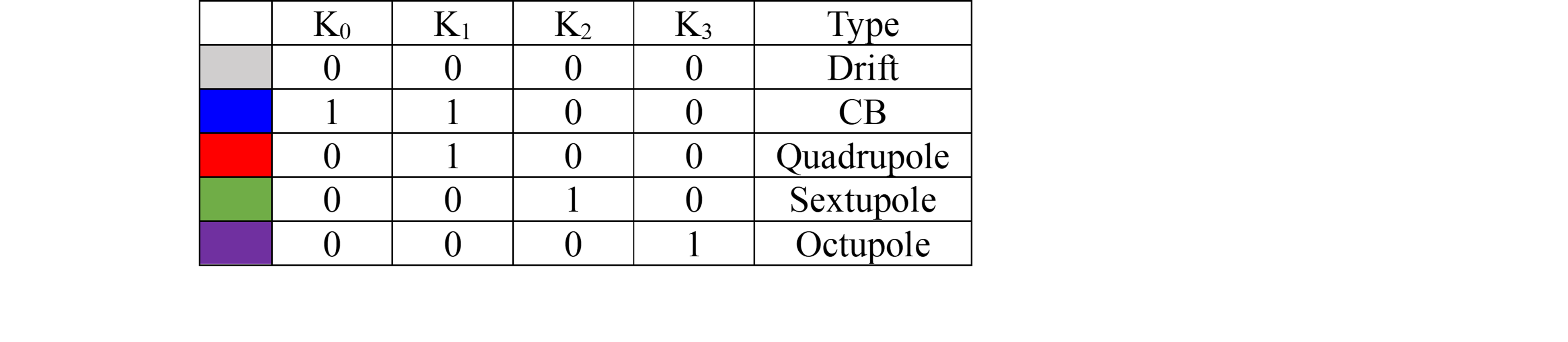}
    \caption{The arrangement of bins for a single cell, where adjacent bins of the same type are grouped into an element. The type of element is marked with a specific color. The two outer CBs are located within the ranges of 5.8 to 7.2~m and 20.5 to 21.9~m. The central CB is located at 10.6 to 17.1~m with a 30-cm gap for a 3PW in the middle. The dispersion bumps are the regions between the central CB and the two outer CBs.}
     \label{table_bins}
\end{figure}

We start our design effort by introducing a method that we call the binning principle in the lattice layout~\cite{Timur_Methodology_TCBA}. This principle standardizes the lattice as a sequence of bins of the following five element types: drift, complex bend (CB), quadrupole, sextupole, and octupole. For instance, according to Fig.~\ref{table_bins},  a 26.4-m-long single cell of NSLS-II upgrade is represented as a sequence of 264 10-cm-long bins, where adjacent bins with the same element type are grouped into an element with each type marked by a different color. It should be noted that the actual layout of elements may be adjusted based on the tunnel geometry and further design optimization. 

The choice of the bin's length as 10~cm comes from the basic scale of the magnetic design of CB PMQs. According to our analysis, the lengths of permanent magnet elements PMQs will be 20~cm or 30~cm, producing a natural scale of focusing length within a cell.

\begin{figure*}[htbp]
    \centering
    \includegraphics[width=\textwidth]{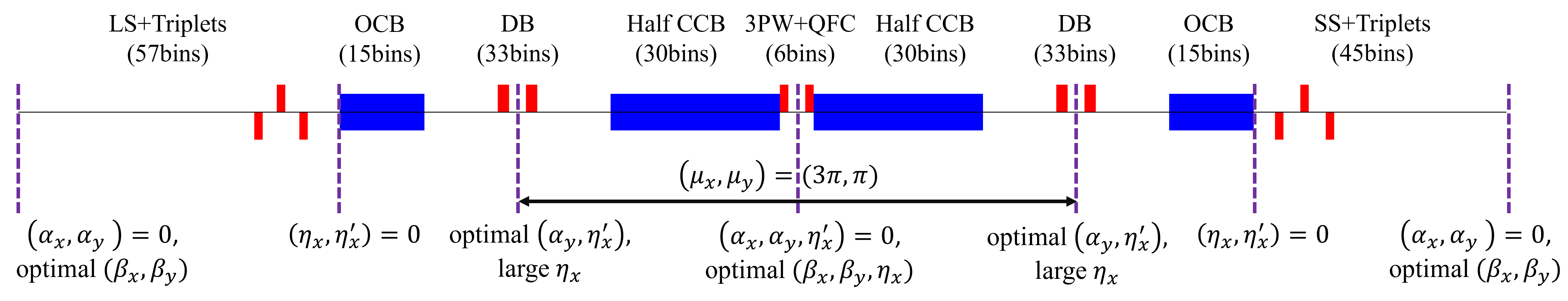}
    \caption{The defined regions of a single cell with their corresponding number of bins, where red and blue blocks represent quadrupoles and complex bends, respectively. The LS, OCB, DB, CCB, 3PW, QFC, and SS stand for long straight, outer complex bend, dispersion bump, central complex bend, 3-pole wiggler, central quadrupole, and short straight\footnote{These abbreviations will be used in the following text.}. We also show that the $\mathcal{M}$-constraints of $\Delta \mathcal{M}$ in the center of long straight, at the exit of outer complex bend, in the center of dispersion bump, in the center of central complex bend and in the center of short straight.}
     \label{layout_regions}
\end{figure*}

Having digitized the lattice into 264 bins of 5 types, we can proceed to optimize the parameters of the bins, focusing on finding the optimal solution for Eqn.~\ref{OP_space} above.
We note that charting the grid of bins on the lattice cell helps to obtain a robust low-emittance solution in an early phase of optimization. Once the solution is demonstrated, we “erase” the grid and continue with the final tuning of the lattice where the locations of components are unconstrained longitudinally unless they are a part of the three complex bends.  

The initial setup of bins contains three element types: drift, quadrupole, and complex bend. Once a reasonable set of linear lattice solutions is established, we switch some of the drift bins to chromatic sextupoles for chromaticity correction, to harmonic sextupoles and octupoles for optimization of dynamic aperture and momentum aperture, and to correctors or BPMs for orbit correction.  

In the next step, the single cell is broken into regions of the bin areas as shown in Fig.~\ref{layout_regions}, which correspond to the layout table presented in Fig.~\ref{table_bins}. Each region is optimized separately in the following order: central CB, dispersion bump, outer CB, and straight sections. The matching conditions (constraints on the Twiss parameters) related to the $\mathcal{M}$-constraints of $\Delta \mathcal{M}$ in Eqn.~\ref{OP_space} need to be satisfied when interfacing these regions:
\begin{itemize}
    \item The mirror condition ($\alpha_x$, $\alpha_y$, $\eta_x'$) = 0 should be matched in the center of central CB, and ($\beta_x$, $\beta_y$, $\eta_x$) could be determined subject to further optimization;
    \item In the center of dispersion bump, a large dispersion height $\eta_x$ is desired. In addition, optimal $\eta_x'$ and $\alpha_y$ are set to create a dispersion region for placing sextupoles;
    \item The phase advance between two dispersion bumps should be matched close to ($3\pi$, $\pi$) so that the $-\mathcal{I}$ transformation between a pair of chromatic sextupoles will help to cancel third-order geometrical RDTs within a cell;
    \item At the exit of outer CB, the $\eta_x$ and $\eta_x'$ are matched to zero for dispersion-free straight sections;
    \item In the center of straights/IDs, the Twiss functions ($\beta_x$, $\beta_y$) are matched to optimum values for higher brightness of the photon beam with ($\alpha_x$, $\alpha_y$) = 0 for creating the beam waists;
    \item The overall beta functions do not exceed 25~m to keep chromaticities within a reasonable range;
    \item Length for each region is fixed as defined by the geometry of the tunnel.
\end{itemize}

In addition, the $P$-constraints of $\Delta s$ and $\Delta p$ in Eqn.~\ref{OP_space} include:
\begin{itemize}
     \item The bending angles for central CB and outer CB are chosen as 4 degrees and 2 degrees, respectively;
     \item The bending field less equal than 0.5~T (i.e., bending angle $K_0\leq$15~mrad), and gradient value $K_1\leq$130~T/m as imposed by the field limitations at the chosen PMQ bore of 16 mm.
\end{itemize} 

At this point, the $P$-space is defined.
It is worth noting here that the bins could be redefined during the optimization process. For instance, one might want to increase an ID straight or have to find a better match for adjusting the lattice to the NSLS-II floor coordinates. It is relatively easy to accomplish, as sliding the bins in either direction and changing their properties is realized in a 1D vector describing the whole cell. 
Furthermore, this technique makes it seamless to develop special cells, such as injection or double minimum beta cells, as just a few bins are re-optimized. Lastly, if we need a lower emittance lattice, it is straightforward to change several bins from drifts to CB PMQs and rerun the optimization loop.

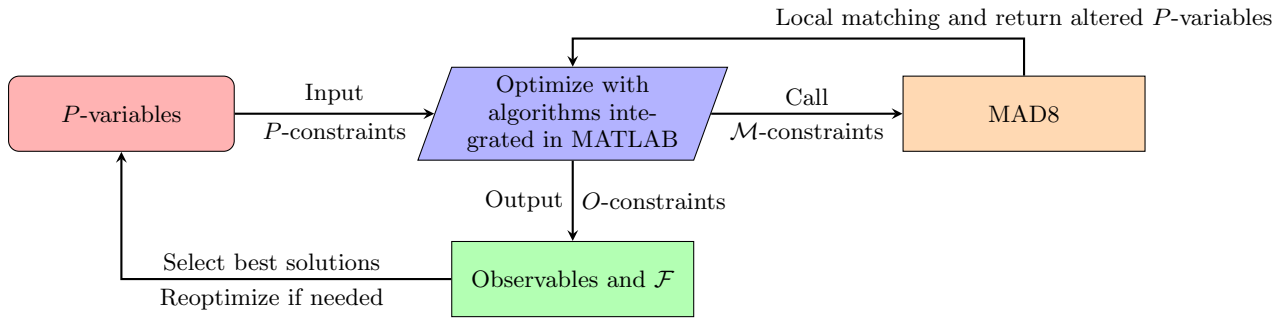
\begin{figure*}[htb]
\tikzstyle{P-var} = [rectangle, rounded corners, 
minimum width=3cm, 
minimum height=1cm,
text centered, 
draw=black, 
fill=red!30]

\tikzstyle{MATLAB} = [trapezium, 
trapezium stretches=true, 
trapezium left angle=70, 
trapezium right angle=110, 
minimum width=3cm, 
minimum height=1cm, 
text centered,
text width=3cm, 
draw=black, fill=blue!30]

\tikzstyle{MAD8} = [rectangle, 
minimum width=3cm, 
minimum height=1cm, 
text centered, 
text width=3cm, 
draw=black, 
fill=orange!30]

\tikzstyle{Observable} = [rectangle, 
minimum width=3cm, 
minimum height=1cm, 
text centered,
text width=3cm, 
draw=black, 
fill=green!30]
\tikzstyle{arrow} = [thick,->,>=stealth]

\begin{tikzpicture}[node distance=2cm]

\node (P-var) [P-var] {$P$-variables};
\node (MATLAB) [MATLAB, right of=P-var, xshift=4cm] {Optimize with algorithms integrated in MATLAB};
\node (MAD8) [MAD8, right of=MATLAB, xshift=4cm] {MAD8};
\node (Observable) [Observable, below of=MATLAB, yshift=-0.2cm] {Observables and $\mathcal{F}$};

\draw [arrow] (P-var) -- node[anchor=south] {Input} node[anchor=north] {$P$-constraints} (MATLAB);
\draw [arrow] (MATLAB) -- node[anchor=south] {Call} node[anchor=north] {$\mathcal{M}$-constraints} (MAD8);
\draw [arrow] (MAD8) -- +(0,1) node[anchor=south] {Local matching and return altered $P$-variables} -|  (MATLAB);
\draw [arrow] (MATLAB) -- node[anchor=west] {$O$-constraints} node[anchor=east] {Output} (Observable);
\draw [arrow] (Observable) -| node[anchor=north, xshift=2cm] {Reoptimize if needed} node[anchor=south, xshift=2cm] {Select best solutions} (P-var);

\end{tikzpicture}
\caption{Flowchart of optimization process.}
\label{Flowchart_optimization}
\end{figure*}

\subsection{Setup of the O-space}

Next, we discuss the process of setting up the $O$-space. The criteria for a new ring lattice performance include:
\begin{itemize}
    \item[$C_1.$] Low natural emittance $\epsilon_{x0}$ for higher brightness;
    \item[$C_2.$] Low chromaticities $\bm{\xi}$;
    \item[$C_3.$] Reasonable radiation loss per turn $U_0$;
    \item[$C_4.$] Reasonable momentum compaction $\alpha_c$;
    \item[$C_5.$] Long straight sections $\bm{L_s}$ for placing longer IDs and other equipment;
    \item[$C_6.$] Optimal beta functions $\bm{\beta_{ID}}$ in the center of IDs for higher brightness of photon beam.
\end{itemize}
 
These figures of merits define the $O$-space. The selection of the best solution is in finding a lattice yielding a minimum of function $\mathcal{F}$, which is defined as a weighted sum of criteria in the $O$-space:
\begin{equation}
    \mathcal{F}=\sum_i \omega_i C_i,
    \label{figure_of_merit}
\end{equation}
where $C_i$ represents each criterion and $\omega_i$ indicates the weight assigned to each criterion. Inverse for some of the components of the goal function $\mathcal{F}$ is used, for instance, for $1/\alpha_c$ and $1/\bm{L_s}$ as large $\alpha_c$ and $\bm{L_s}$ are desired.

We have chosen the $O$-constraints of $\Delta O$ in the Eqn.~\ref{OP_space} as the following:
\begin{itemize}
    \item Horizontal partition number $J_x$ within the range of [1.8, 2.3];
    \item Momentum compaction $\alpha_c>5\times10^{-5}$;
    \item Minimum lengths $\bm{L_s}$ of long straight and short straight are 8.4~m and 5.4~m, respectively;
    \item The Twiss functions $\bm{\beta_{ID}}$ in the center of IDs are less than 3~m.
\end{itemize}

\subsection{Setup of the optimization method}\label{optimizaton_method}

Figure~\ref{Flowchart_optimization} illustrates the optimization process for solving Eqn.~\ref{OP_space}~\cite{Timur_Methodology_TCBA}. The NSGA-II algorithm~\cite{Deb_NSGAII} integrated into the MATLAB scripts~\cite{matlab_NSGA-II} has been employed with an interface to the MAD8 program~\cite{Iselin_MAD8} for carrying out the local matching. Local matching is realized in MAD8 under constraints given by initial and final Twiss parameters. In NSGA-II, the chromosome is a vector of $P$-variables, and the gene is a variable. The altered $P$-variables (gradient values) are then carried over to the chromosome to generate a new population. Under the $O$-constraints, the optimization outputs the observables and calculates the figure of merit $\mathcal{F}$. This guides us in selecting the best solution or, if necessary, conducting the next round of optimization. For the setup of the NSGA-II algorithm, the simulated binary crossover (SBX) and polynomial mutation (PLM) are used with the distribution indices for crossover and mutation equal to 20. The crossover probability is set to 0.9, and mutation probability is chosen as $1/n$, where $n$ is the number of decision variables. To evenly sample the points in the decision space, the Latin hypercube sampling (LHS)~\cite{LHS} method is used to generate the initial population. The population size in each generation is set to 2000.

\section{Design of CBA lattice}\label{result}

\begin{figure*}[htbp]
    \centering
    \includegraphics[width=\textwidth]{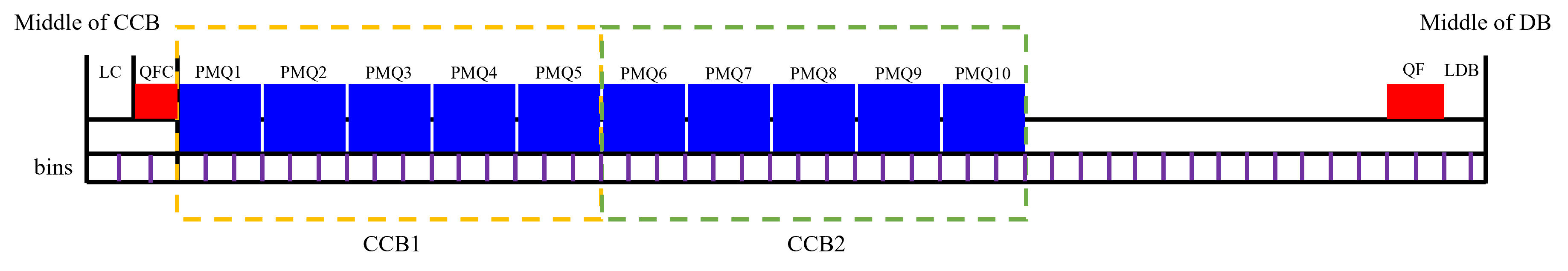}
    \caption{Illustration of layout for half central CB containing 10 PMQs and half dispersion bump. LC is the half drift length of the reserved space for a 3PW. QFC is the center quadrupole that matches mirror conditions in the middle of CCB. QF is the quadrupole placed in the dispersion bump, and LDB is the half drift length between two QFs.}
     \label{HCCB_sketch}
\end{figure*}

The natural emittance, radiation loss, and momentum compaction of a CBA lattice have the following relations with radiation integrals~\cite{Lee_accel_phys_book}:
\begin{equation}
    \epsilon_{x0} \propto \frac{I_5}{I_2-I_4}, U_0 \propto I_2, \alpha_c \propto I_1.
\end{equation}
The evaluations of radiation integrals contributed from short PMQs of complex bends are expressed as\cite{Helm_radiation_integrals}:
\begin{eqnarray}
    I_1&=&\oint \eta_x h ds\approx\sum_i \langle \eta_x \rangle_i{K_{0,i}},\\
    I_2&=&\oint h^2 ds\approx \frac{1}{L}\sum_i {K_{0,i}^2},\\
    I_3&=&\oint |h|^3 ds\approx \frac{1}{L^2}\sum_i {|K_{0,i}|^3},\\
    I_4&=&\oint \eta_x(h^3+2hK_1)ds\approx\sum_i (\frac{K_{0,i}^3}{L^2}+2K_{0,i} K_{1,i})\langle \eta_x \rangle_i,\\
    I_5&=&\oint |h|^3 \mathcal{H}ds\approx \frac{1}{L^2}\sum_i {|K_{0,i}|^3}\langle \mathcal{H} \rangle_i,\label{I5}
\end{eqnarray}
where $h=\frac{1}{\rho}$ is the orbit curvature, $\mathcal{H}=\gamma_x\eta_x^2+2\alpha_x\eta_x\eta_x'+\beta_x\eta_x'^2$ is the Courant-Snyder invariant defined by Twiss parameters ($\alpha_x$, $\beta_x$, $\gamma_x$) and dispersion functions ($\eta_x$, $\eta_x'$). By taking advantage of the binning method mentioned in the last section, the radiation integrals are estimated via summations through the complex bend regions, where $K_{0,i}$ and $K_{1,i}$ are the bending angle and gradient value for $i$th PMQ with a constant length of $L$ = 3bins. In these equations, $\langle \eta_x \rangle_i$ and $\langle \mathcal{H} \rangle_i$ correspond to $\eta_x$ and $\mathcal{H}$ of $i$th PMQ averaged along the respective bins.

In addition, the natural chromaticities are estimated by\cite{Lee_accel_phys_book}:
\begin{equation}
    \bm{\xi}=\frac{-1}{4\pi}\oint \bm{\beta}K_1ds \approx \frac{-L}{4\pi} \sum_i \langle \bm{\beta} \rangle_i {K_{1,i}},
\end{equation}
where $\langle \bm{\beta} \rangle_i$ denotes to the averaged $\bm{\beta}=(\beta_x, \beta_y)$ of $i$th PMQ.

Therefore, to obtain the best $\mathcal{F}$ in terms of the criteria considered in Eqn.~\ref{figure_of_merit}, the $P$-variables $K_{0,i}$ and $K_{1,i}$ of the central and outer complex bends need to be optimized. 

\subsection{{Optimization of central complex bend and dispersion bump}}

Figure~\ref{HCCB_sketch} displays the layout of half central CB and half dispersion bump. In the middle of CCB, a drift space is reserved for a source of 3PW beamline, and a center quadrupole is also used to match mirror conditions. In the dispersion bump region, two symmetrical QFs are separated with a drift space for placing a sextupole. Furthermore, the CCB1 (PMQ1 to PMQ5) and CCB2 (PMQ6 to PMQ10) are optimized separately.

To optimize the central complex bend, we define $K_{0,i}$ and $K_{1,i}$ of PMQ1 to PMQ10 as the following distributions:

\begin{eqnarray}
      K_{0,i} &=&
    \begin{cases}
      A_1, & i=1,3,5\\
      A_2, & i=2,4\\
      \sum_m a_m(i-6)^m, & i=6,...,10, m=0,..,3
    \end{cases}  \\
  K_{1,i} &=&
    \begin{cases}
      B_1, & i=1,3,5\\
      B_2, & i=2,4\\
      \text{optimal} \ K_{1,i}, & i=6,...,10
    \end{cases} 
\end{eqnarray}
where $i$ is the index of PMQ and $a_m$ are the coefficients of a polynomial function. In addition, $A_1$, $A_2$, $B_1$, and $B_2$ are the coefficients of step-wise periodic functions. The conversion of optimization variables to the coefficients of functions reduces the number of $P$-variables, which significantly benefits the search for optimal solutions in a complicated decision space. 

\begin{figure}[htbp]
    \centering
    \includegraphics[width=\columnwidth]{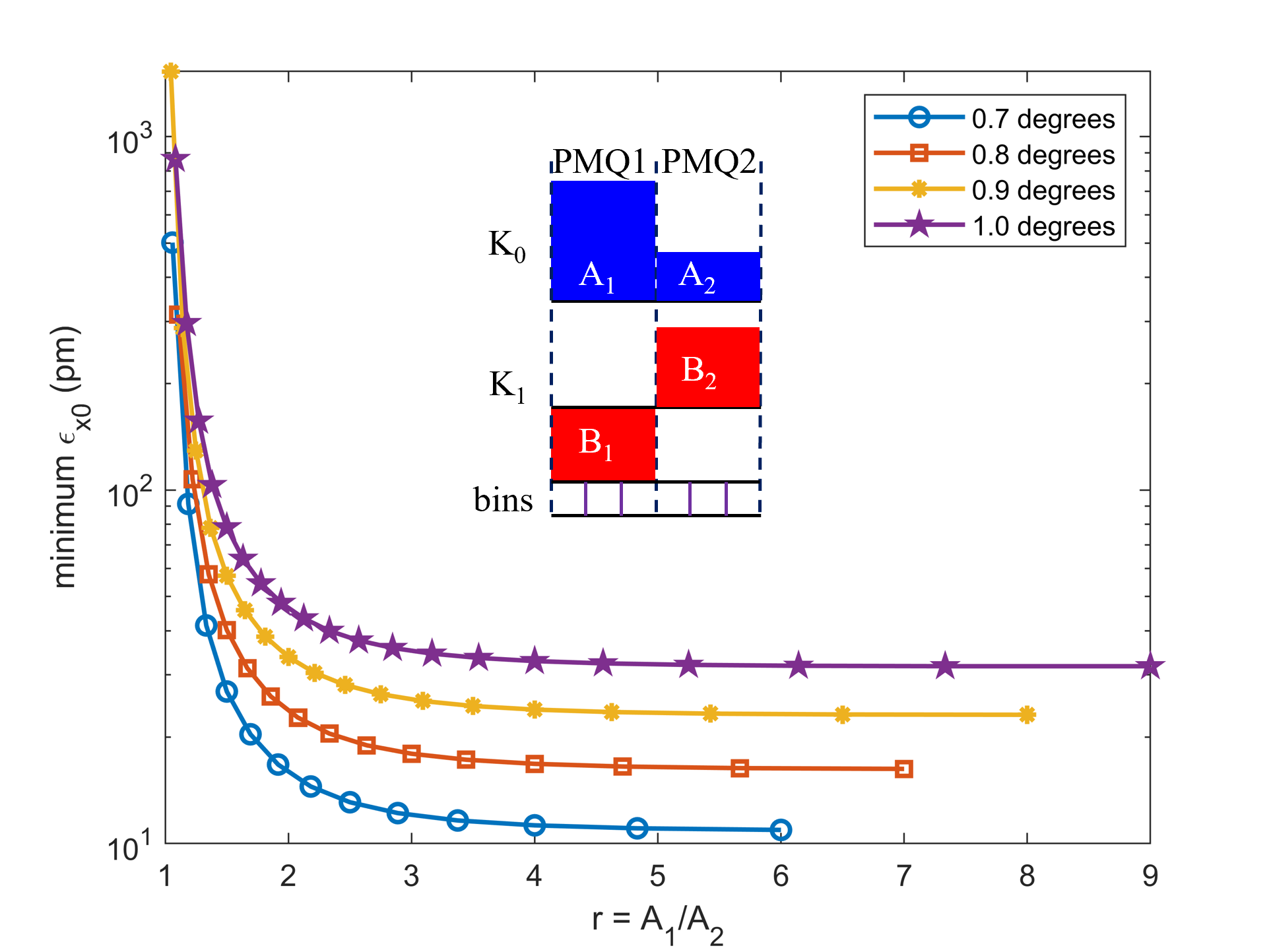}
    \caption{The scan results for FODO-like PMQs (PMQ1 and PMQ2) with four different sum bending angles ($A_{sum}=A_1+A_2$) range from 0.7 to 1.0 degrees with an interval of 0.1 degrees. The plot displays the minimum natural emittance versus the ratios of bending angles ($r=A_1/A_2$). The inner picture indicates that PMQ with a large bending angle should be defocusing.}
     \label{fodo_scan}
\end{figure}

The optimal $A_1$, $A_2$, $B_1$, and $B_2$ for CCB1 are found by scanning FODO-like PMQs (PMQ1 and PMQ2) with the objective of minimum natural emittance. Figure~\ref{fodo_scan} shows the scan results for different $r$ for different $A_{sum}$. With the considerations of natural emittance and allocation of bending angles for CCB1 and CCB2, we choose $A_1$ = 0.6 degrees and $A_2$ = 0.2 degrees with their corresponding gradient values are $B_1$ = -110~T/m and $B_2$ = 130~T/m, respectively.  

\begin{figure}[htbp]
    \centering
        \includegraphics[width=\columnwidth]{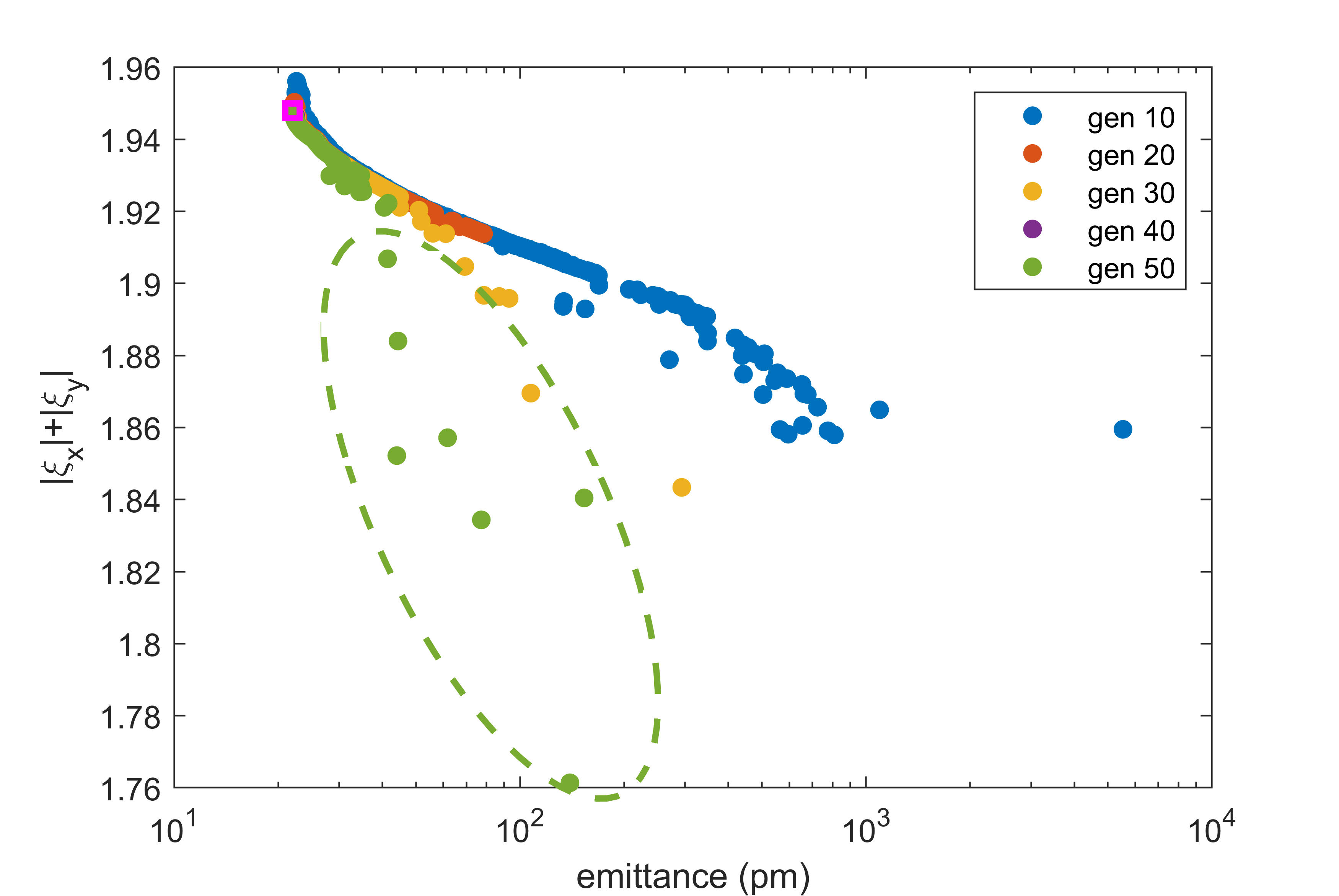}
    \caption{Evolution of distributions of objective values in generations 10, 20, 30, 40, and 50, where different colors represent different generations and the selected solution is marked with a magenta box.}
     \label{HCCB_DB_NSGAII}
\end{figure}

We optimize CCB2 and half dispersion bump simultaneously using the method described in Section~\ref{optimizaton_method}. The optimization goal is to minimize natural emittance $\epsilon_{x0}$ and sum of absolute natural chromaticities $|\xi_x|+|\xi_y|$ simultaneously. The optimization variables include $a_m$, $K_{1,i}$ of CCB2, $K_1$ of QF and drift length LDB. The optimization objectives are evaluated from the middle of CCB to the middle of DB. Figure~\ref{HCCB_DB_NSGAII} presents the evolution of distributions of objective values from generation 10 to generation 50 with an interval of 10 generations. The optimization is almost converged after 50 generations. The overall pattern of distributions of objective values in generation 50, except for a few points in the dashed ellipse, forms the shape of a ``short island". The selected solution is marked with a magenta box with an emittance value of about 22~pm. 

\begin{figure}[htbp]
    \centering
        \includegraphics[width=\columnwidth]{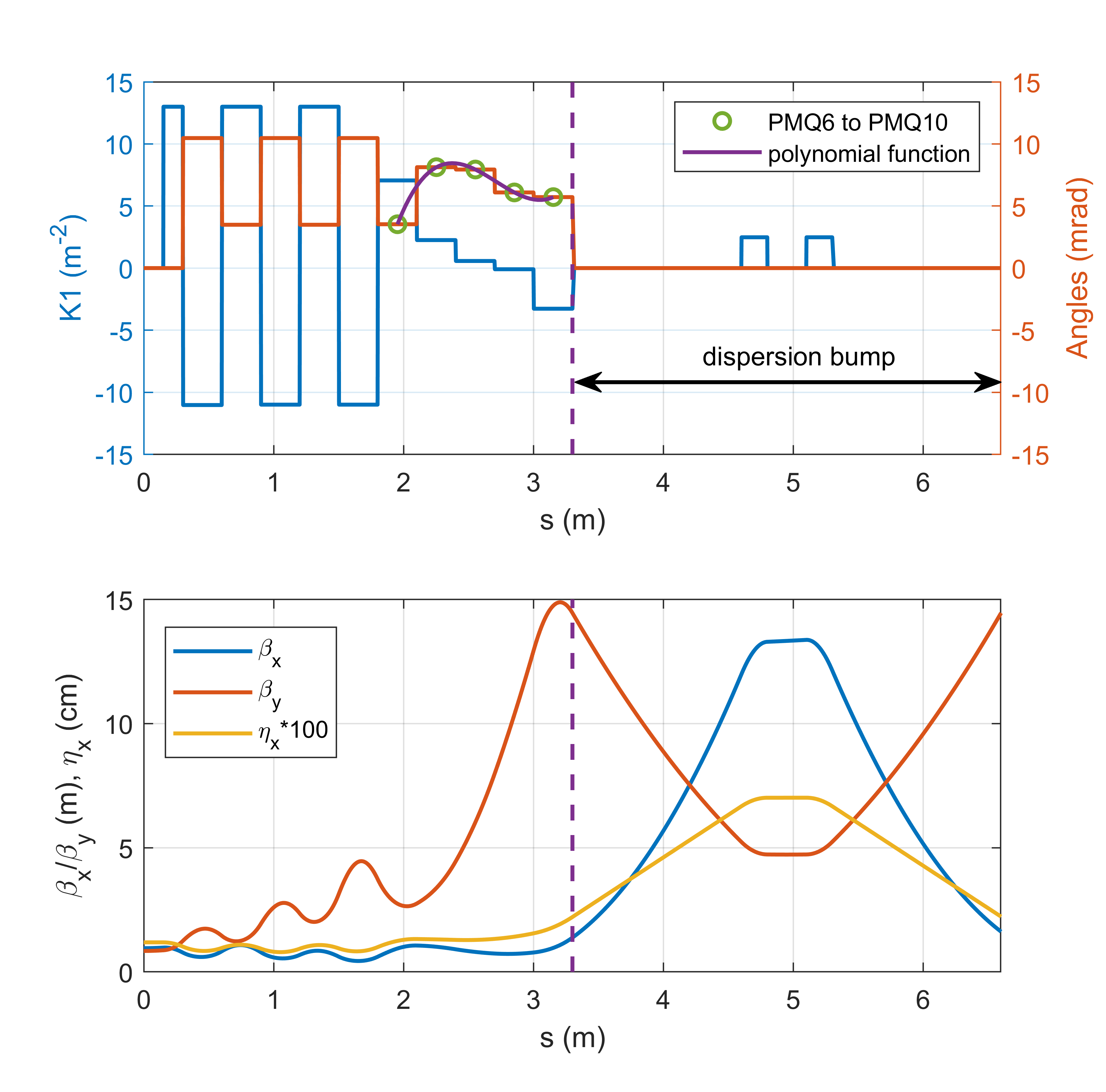}
    \caption{Top plot: Field values from the middle of CCB to the middle of DB, including the values of field gradients and bending angles, where green circles show the selected bending angles of CCB2 (PMQ6 to PMQ10) follow a polynomial function; Bottom plot: The corresponding Twiss function values from the middle of CCB to the end of DB.}
     \label{optimized_HCCB_DB}
\end{figure}

Figure~\ref{optimized_HCCB_DB} presents the field values and corresponding Twiss function values for the selected solution, where the purple dashed line indicates the end of half CCB and the start of the dispersion bump. A short 3PW is planned to be installed in the gap within the central complex bend for the source of a beamline and aligned to the window in the rachet wall. In the top plot, the optimized bending angles of CCB2 are represented by the green circles, which show they follow a polynomial function. The bottom plot shows Twiss function values from the middle of CCB to the end of the dispersion bump, and the phase advance is matched close to (1.5$\pi$, 0.5$\pi$). Therefore, when introducing three sextupole families (one for focusing and two for defocusing) in the two dispersion bumps, the symmetrical beta functions ($\beta_x$, $\beta_y$) and the phase advance of (3$\pi$, $\pi$) between three pairs of sextupoles will enable effective cancellation of third order geometrical resonance driving terms (RDTs) within a cell.  

\subsection{{Optimization of outer complex bend}}

\begin{figure}[htbp]
    \centering
    \includegraphics[width=0.9\columnwidth]{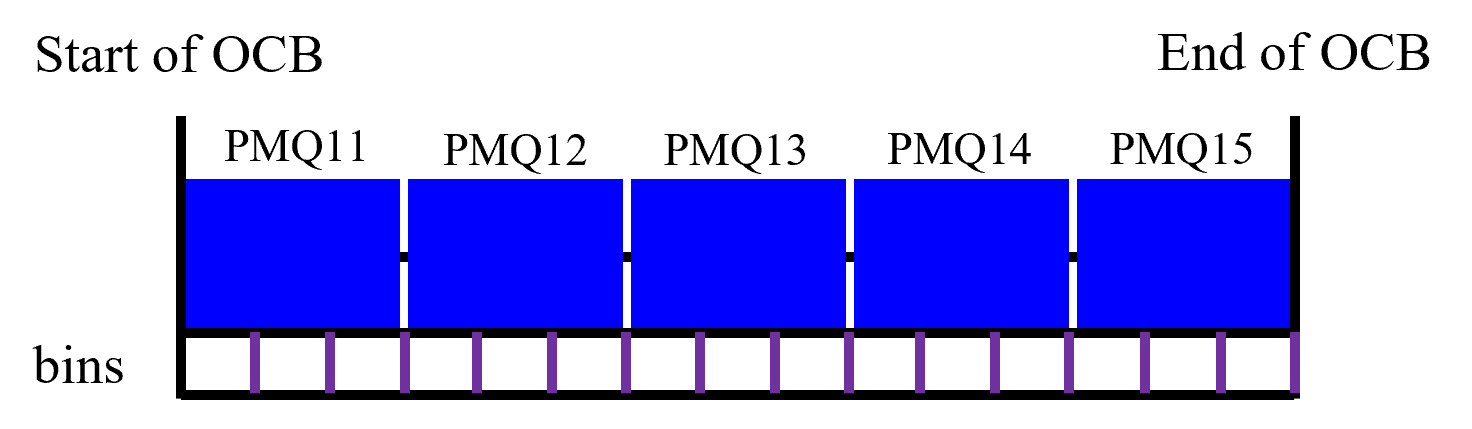}
    \caption{Illustration of layout for OCB containing PMQ11 to PMQ15.}
     \label{OCB_skectch}
\end{figure}

\begin{figure}[htbp]
    \centering
        \includegraphics[width=\columnwidth]{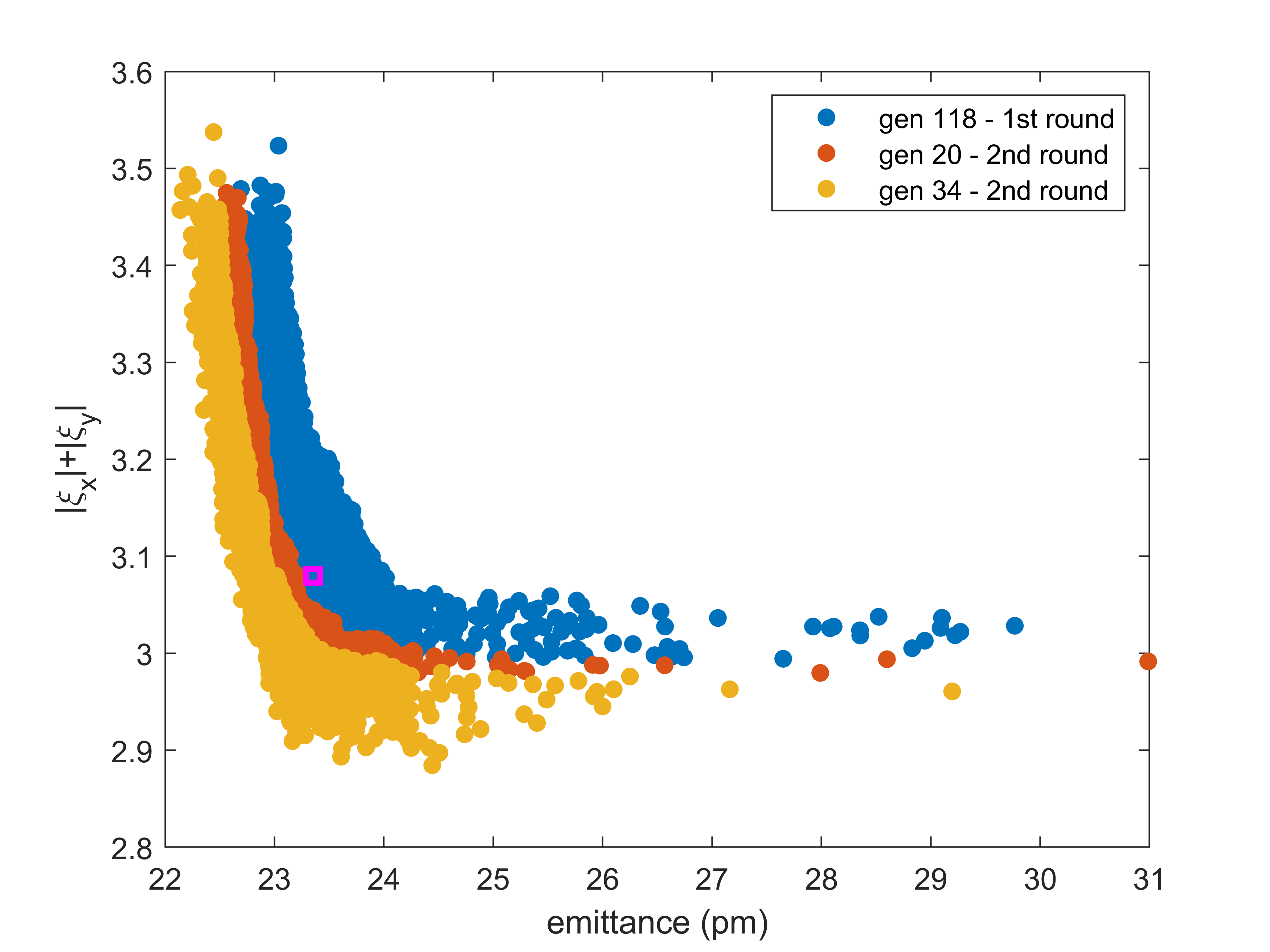}
    \caption{Evolution of distributions of objective values in generation 118 of the first round and in generations 20 and 34 of the second round, where different colors represent different generations, and the selected solution is marked with a magenta box.}
     \label{OCB_NSGAII}
\end{figure}

Figure~\ref{OCB_skectch} illustrates the layout of OCB containing PMQ11 to PMQ15. We use similar variables and objectives for OCB optimization as we did for CCB2 optimization. The objectives are evaluated from the middle of CCB to the end of OCB. Figure~\ref{OCB_NSGAII} shows the optimization results with two rounds using the method described in Section~\ref{optimizaton_method}. The second round can start from the first round due to the stochastic characters of the NSGA-II algorithm. As shown in Fig.~\ref{OCB_NSGAII}, the optimization almost converges after 34 generations in the second round. It is clear to note the "chromaticity wall" when pushing emittance close to 22~pm. This shape of distributions of objective values indicates that a solution with lower chromaticities can be selected without sacrificing emittance too much. The selected solution is marked with a magenta box in Fig.~\ref{OCB_NSGAII}, which indicates the ring's natural emittance is about 23~pm.    

\begin{figure}[htbp]
    \centering
        \includegraphics[width=\columnwidth]{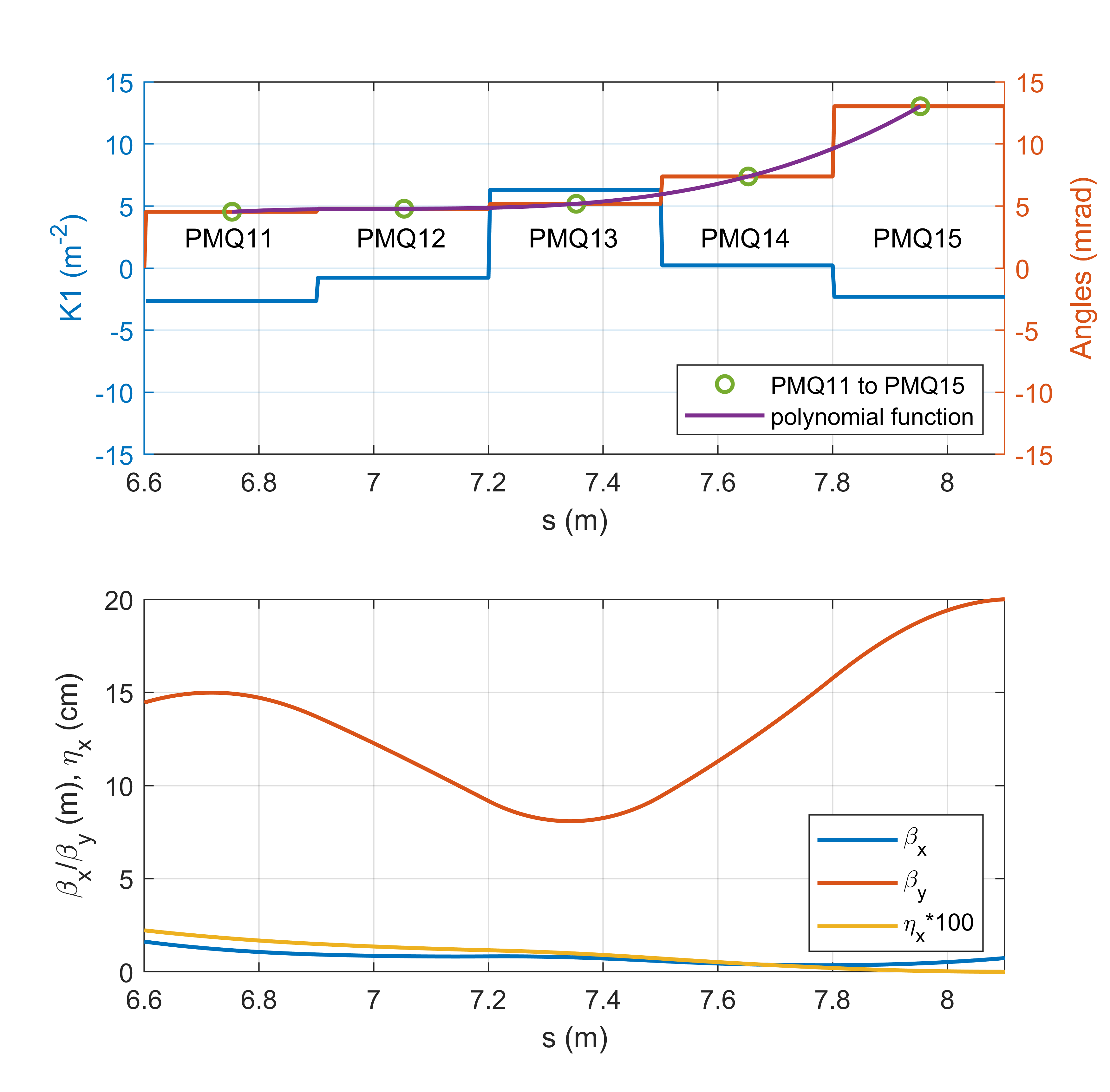}
    \caption{Top plot: Field values from the start of OCB to the end of OCB including the values of field gradients and bending angles, where green circles show the selected bending angles of PMQ11 to PMQ15 follow a polynomial function; Bottom plot: The corresponding Twiss function values from the start of OCB to the end of OCB.}
     \label{optimized_OCB}
\end{figure}

The top plot of Fig.~\ref{optimized_OCB} shows the values for the magnetic fields and field gradients for the selected solution for the outer complex bend, where the green circles represent bending angles of PMQ11 to PMQ15. The bottom plot of Fig.~\ref{optimized_OCB} presents the Twiss function values along the outer complex bend. 

\subsection{Straight section design}

As illustrated in Fig.~\ref{layout_regions}, the long straight section and matching triplets are proposed to have 57 bins, while the short straight section and matching triplets are defined to have 45 bins. If referring to the layout of the NSLS-II straight sections for diagnostic instrumentation, vacuum components, RF cavity, etc., the $O$-constraints of required minimum lengths for long straight and short straight should be 8.4~m and 5.4~m, respectively. Given that the initial Twiss function values, the $P$-variables of positions and strengths of matching triplets are optimized with the $P$-constraints to achieve the optimal $\bm{\beta}_{ID}$ in the center of straights. 

\subsection{The developed lattice solution}

\begin{figure*}[htbp]
  \includegraphics[width=0.95\textwidth]{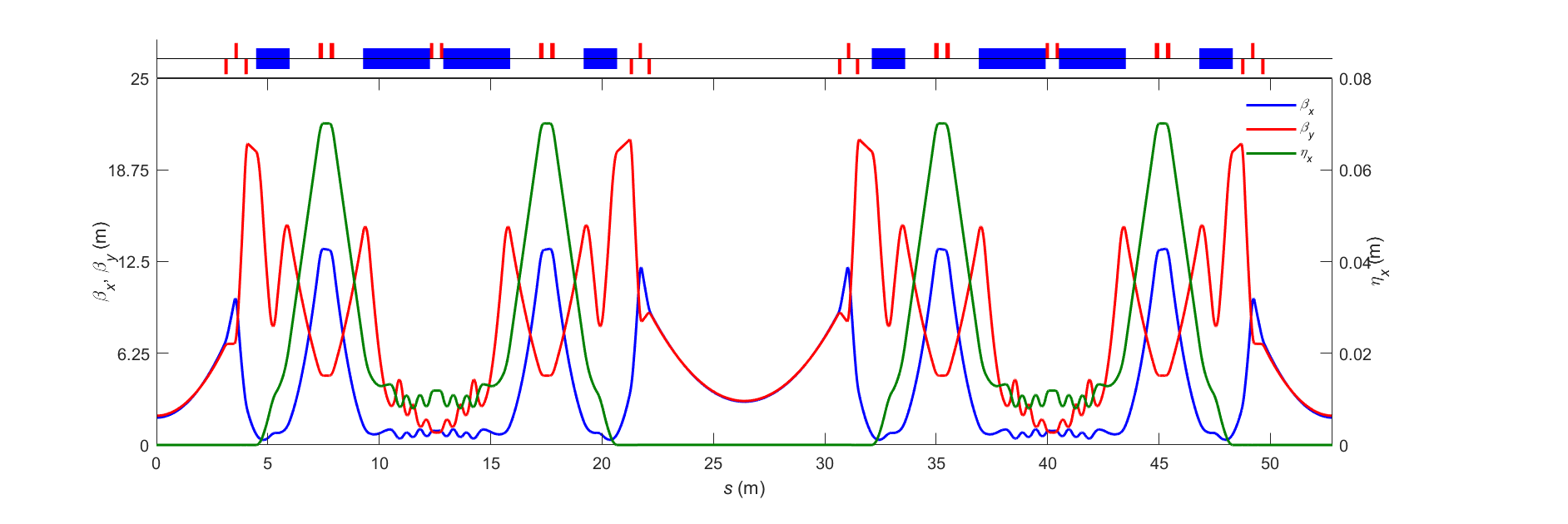}\\
    \includegraphics[width=0.95\textwidth]{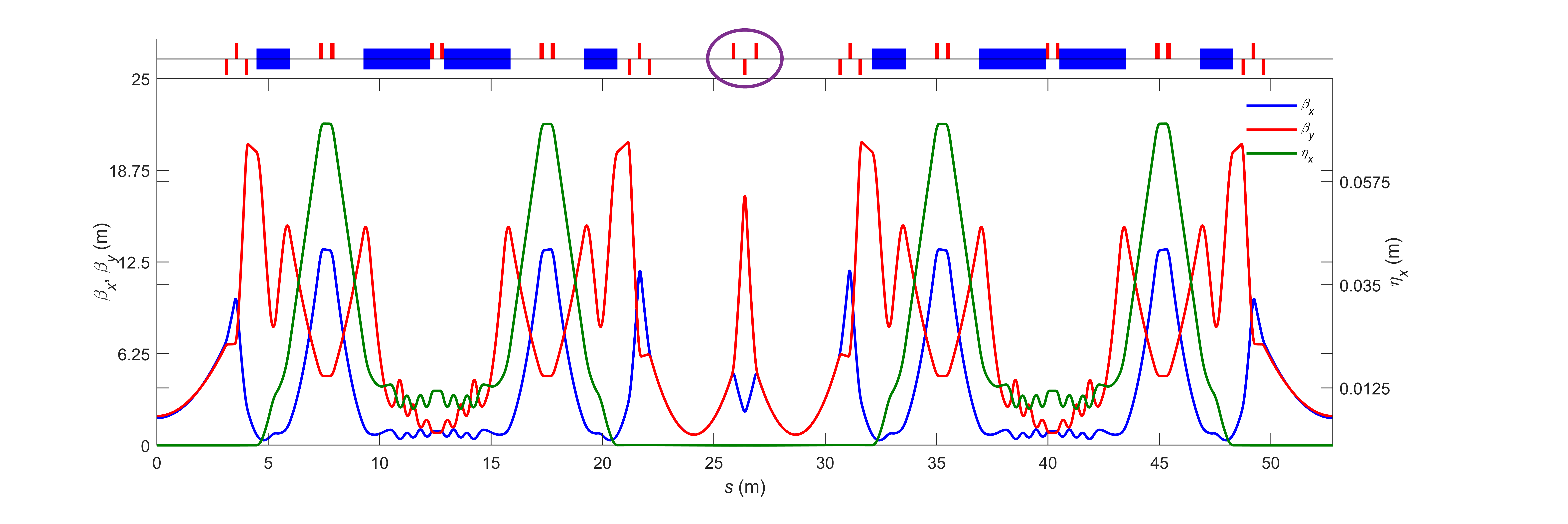}\\
      \includegraphics[width=0.95\textwidth]{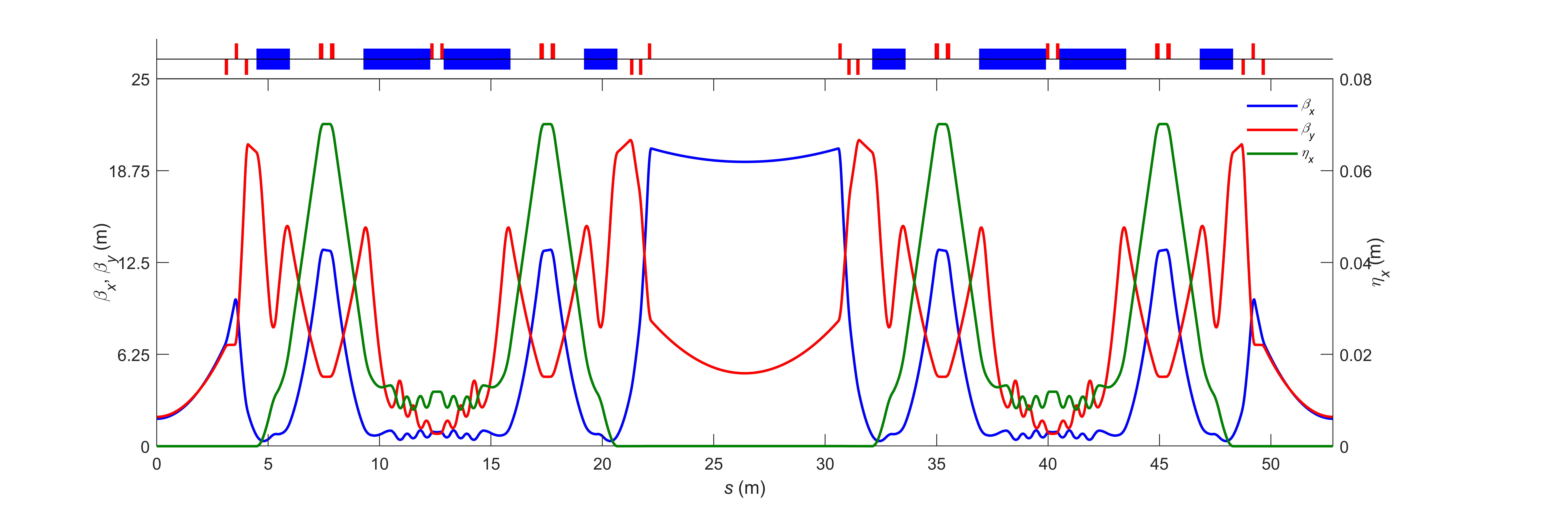}
 \caption{The Layout of magnets and Twiss function values along one supercell for periodic structure, double minimum beta structure, and high beta structure from the top to the bottom plots, starting from the center of a single short straight to the center of the next short straight. The beam energy is 3~GeV. The complex bends are colored blue, and the red blocks represent the quadrupoles. The sextupoles and octupoles are not shown in this figure.}
 \label{supercell_3layouts}
\end{figure*}

As a result of the optimization described above, we continue with the lattice design by integrating the following regions: central complex bend, dispersion bumps, outer complex bends, and straight sections. The top plot of Fig.~\ref{supercell_3layouts} shows the layout of magnets and Twiss function values along one periodic supercell, starting from the center of a single short straight section to the center of the next short straight section. Due to the compact arrangement of complex bend elements, a substantial amount of free space ($\sim$ 50~\%) is available for various accelerator equipment, including correctors, vacuum, and diagnostic devices. Furthermore, due to our choice of permanent magnet technology, the number of power supplies will be largely reduced, and the power consumption regarding magnets will be roughly 20~\% of that for the NSLS-II lattice. 

\begin{table*}[htbp]
\centering
\caption{The parameters of the NSLS-II lattice and the developed NSLS-IIU CBA lattices comprise 15 periodic supercells operating at 3 GeV and 4 GeV. The LS and SS denote the long straight and short straight, respectively.}
\label{periodic_ring_parameter_table}
\begin{tabular}{c |c |c |c}
\hline\hline
\multicolumn{1}{c|}{Paramters} &
\multicolumn{3}{c}{Values} \\
\hline
\multicolumn{1}{c|}{} &
\multicolumn{1}{c|}{NSLS-II bare lattice} &
\multicolumn{2}{c}{NSLS-IIU CBA lattice} 
\\
\hline
Circumference $C$ [m] & 791.958 & 791.7679 & 791.7522\\
Beam energy $E$ [GeV] & 3 & 3 & 4 \\
Natural emittance $\epsilon_{x0}$ [pm-rad] & 2086 & 23.4 & 43.2 \\
Damping partitions ($J_x, J_y, J_\delta$) & (1, 1, 2) & (2.24, 1, 0.76) & (2.27, 1, 0.73)  \\
Ring tunes ($\nu_x,\nu_y$)  & (33.22, 16.26) & (84.67, 28.87) & (83.17, 29.27) \\
Natural chromaticities ($\xi_x, \xi_y$) & (-98.5, -40.2) & (-135, -144) & (-136, -139) \\
Momentum compaction $\alpha_c$ & 3.63$\times 10^{-4}$ & 7.76$\times 10^{-5}$ & 7.60$\times 10^{-5}$\\
Energy loss per turn $U_0$ [keV] & 286.4 & 196 & 603\\
Energy spread $\sigma_\delta$ [\%] & 0.0514 & 0.073 & 0.097\\
($\beta_x$, $\beta_y$) at LS center [m] & (20.1, 3.4) & (2.95, 2.99) & (2.79, 2.80)\\
($\beta_x$, $\beta_y$) at SS center [m] & (1.8, 1.1) & (1.87, 1.99) & (1.79, 1.81)\\
($\beta_{x,max}$, $\beta_{y,max}$) [m] & (29.99, 27.31) & (13.37, 20.82) & (14.92, 23.71) \\
($\beta_{x,min}$, $\beta_{y,min}$) [m] & (1.84, 1.17) & (0.35, 0.84) & (0.51, 0.80)\\
($\beta_{x,avg}$, $\beta_{y,avg}$) [m] & (12.58, 13.79) & (3.99, 7.51) & (4.61, 7.69)\\
Length of LS [m] & 9.3 & 8.4 & 8.4\\
Length of SS [m] & 6.6 & 6.1 & 6.1\\
Sum of $K_2$ of chromatic sextupoles per cell [m$^{-3}$] & $\sim$100 & $\sim$1600 & $\sim$1600 \\
\hline
\end{tabular}
\end{table*}

The parameters and their values of developed complex bend lattices at 3 and 4 GeV composed from 15 periodic supercells are listed in Table~\ref{periodic_ring_parameter_table} compared to the NSLS-II lattice. The natural emittance of the NSLS-II upgrade lattice is minimized to about 23~pm at 3~GeV, which is about 90 times smaller than the NSLS-II lattice. The beta functions in the center of the long straight section are between 2 and 3~m, while they are between 1 and 2~m in the center of the short straight section. In addition, compared to other MBA lattices~\cite{Fornek_APS-U, Hellert_ALS-U, Jiao_HEPS, Raimondi_ESRF-EBS}, the lengths of straight sections are longer so that long or tandem/canted undulators could be installed. These benefits for user programs demonstrate the advantages of employing complex bend lattice for NSLS-II upgrade.

The middle plot of Fig.~\ref{supercell_3layouts} shows the Twiss function values along one supercell with double minimum beta functions in the long straight section, which are created by putting a triplets in the middle of the long straight section as denoted by the purple ellipse.  

In practice, the NSLS-II upgrade lattice may consist of one supercell with high beta functions shown in the bottom plot of Fig.~\ref{supercell_3layouts} for off-axis injection, and the rest supercells are periodic structures for IDs. Double minimum beta structures will possibly replace several supercells. 

\subsection{Nonlinear dynamics optimization}

\begin{figure}[htbp]
    \centering
    \includegraphics[width=\columnwidth]{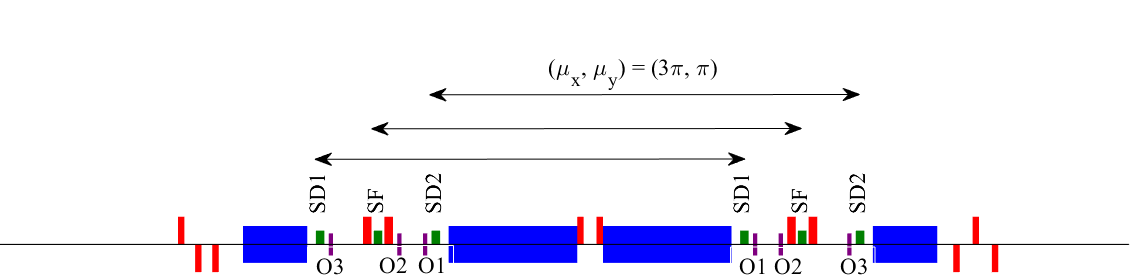}
    \caption{Locations of three chromatic sextupole families and three chromatic octupole families in the high beta cell. The phase advances between three pairs of sextupoles are close to ($3\pi$,$\pi$).}
     \label{layout_sext_oct}
\end{figure}

The nonlinear dynamics optimization targeting sufficient dynamic and momentum apertures is in progress at the time of writing this paper, so we are presenting here some preliminary results. We installed three chromatic sextupole families used to correct the chromaticities ($\xi_x, \xi_y$) to (+2, +2) as shown in Fig.~\ref{layout_sext_oct}. The phase advances between three pairs of sextupoles are close to ($3\pi$,$\pi$) to cancel the third-order geometrical resonance driving terms (RDTs). To further improve the performance of DA and MA, three chromatic octupole families are also placed in phase with three chromatic sextupole families to simultaneously correct the fourth order geometrical resonance driving terms (RDTs) and amplitude-dependent tune shift (ADTs)~\cite{Fabien_octupole_correction}. 

\begin{figure}[htbp]
    \centering
        \includegraphics[width=\columnwidth]{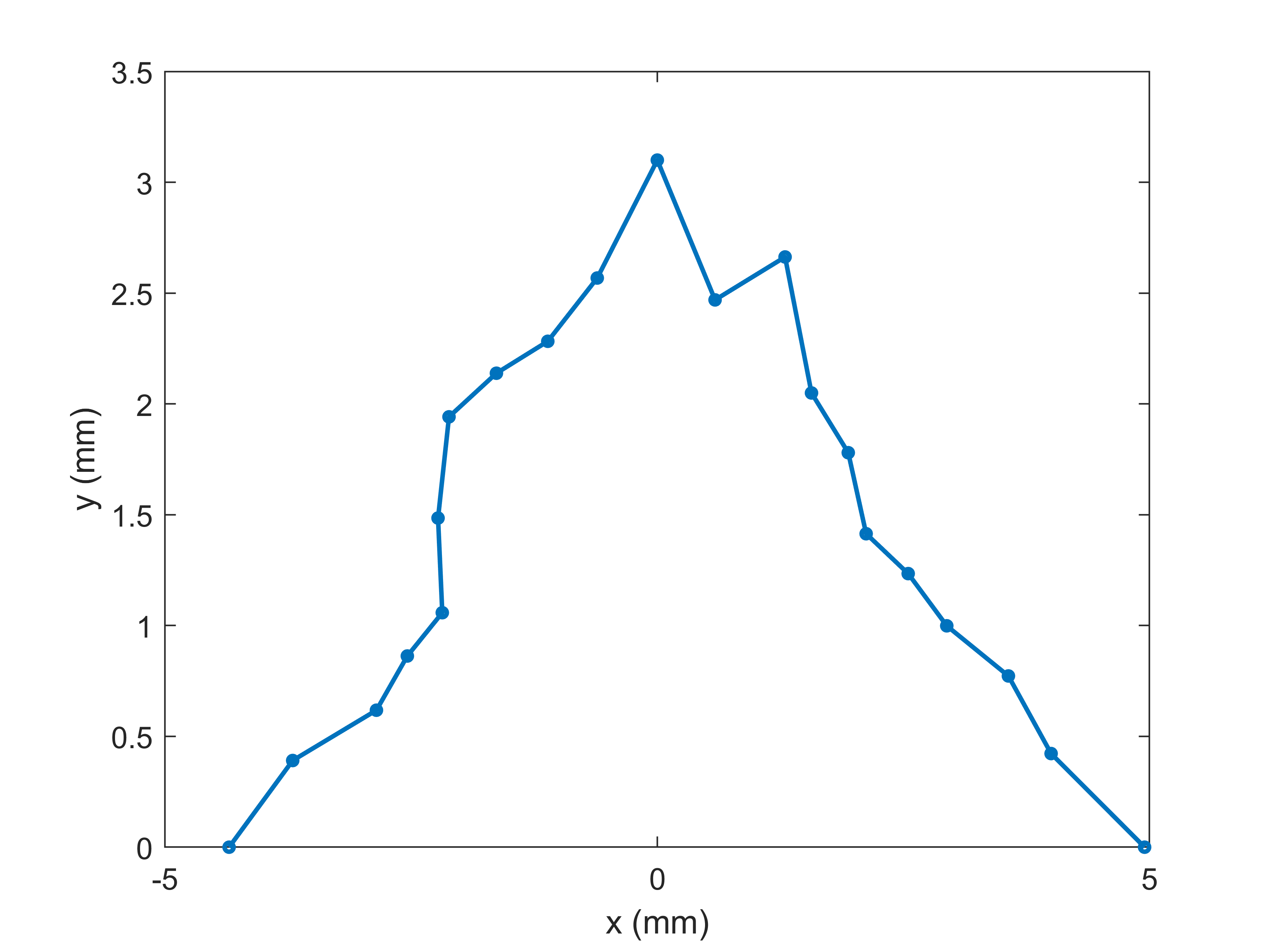}
        \includegraphics[width=\columnwidth]{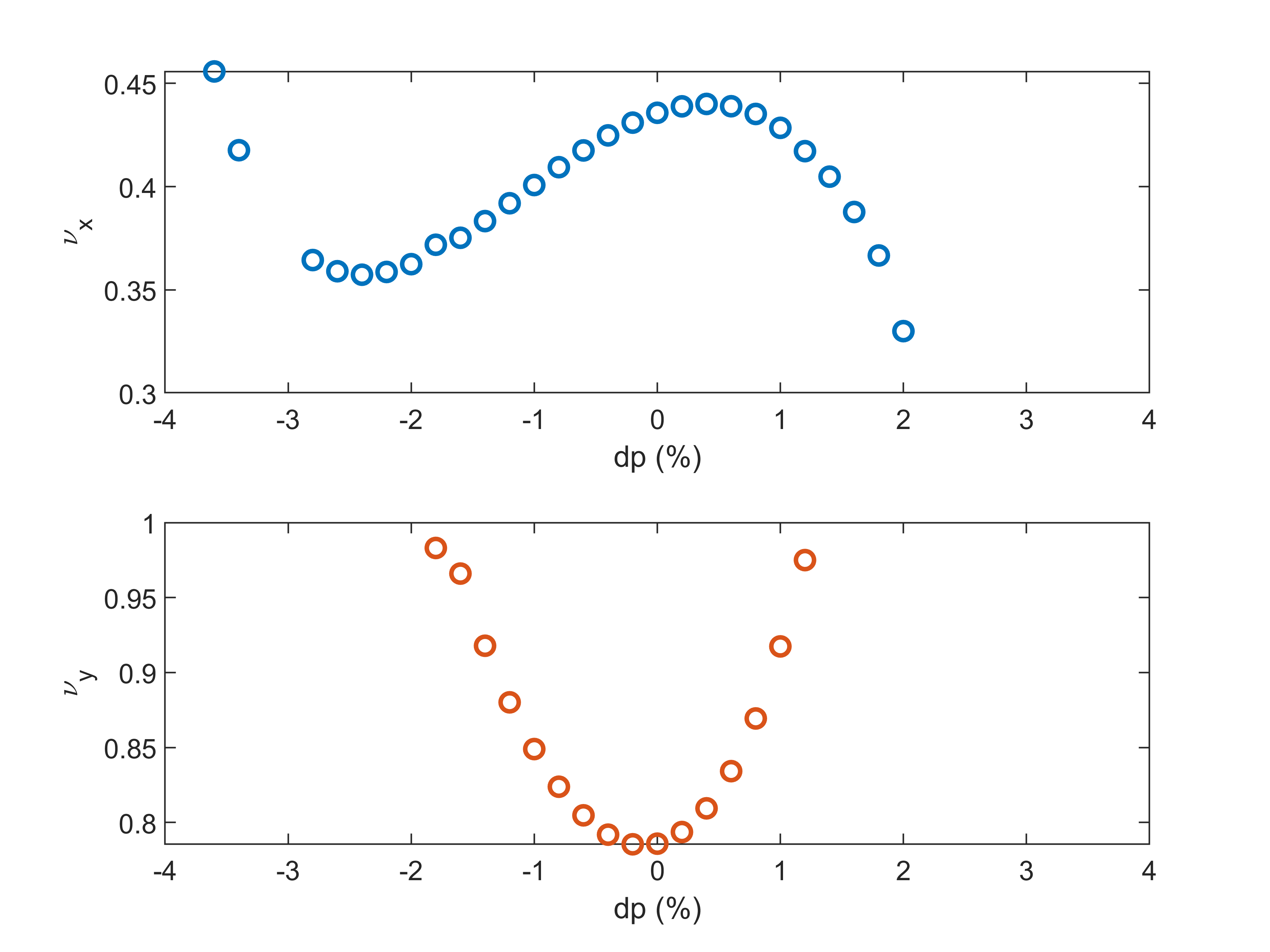}
    \caption{The dynamic aperture (top) and tune shift with momentum (bottom).}
     \label{optimized_DA_MA}
\end{figure}

During the early stage of nonlinear dynamics optimization, we assume that the CB lattice consists of one supercell with a high beta structure and 14 supercells with periodic structures. 
Our preliminary results from optimizing dynamic aperture and tune shift with momentum are shown in Fig.~\ref{optimized_DA_MA}. The achieved range of DA is approximately -5/+5~mm, and MA has reached about 2~\%. To further improve the DA and MA, the ring tunes are scanned, and the positions of chromatic sextupoles and chromatic octupoles will be optimized.
In addition, the harmonic sextupoles will also be used to reduce tune shift with amplitude to maximize the dynamic aperture~\cite{Minghao_retreat_slides}.

Table~\ref{Compare_MBA&CBA} shows a comparison of multi-bend lattices of several facilities with the complex bend lattice for NSLS-IIU. 

\begin{table*}[htbp]
\centering
\caption{Comparison of multi-bend lattices from several facilities with respect to the complex bend lattice for NSLS-IIU.}
\label{Compare_MBA&CBA}
\begin{tabular}{c |c c c c c c c| c}
\hline\hline
& APS-U~\cite{Fornek_APS-U} & SIRIUS~\cite{Rodrigues_Sirius} & ALS-U~\cite{Hellert_ALS-U} & SLS-II~\cite{Streun_SLSII} & ESRF-EBS~\cite{Raimondi_ESRF-EBS} & Korea-4GSR~\cite{Jang_Korea-4GSR} & HEPS~\cite{Jiao_HEPS} & NSLS-IIU CBA\\
\hline
$C$ [m] & 1103.61 & 518.4 & 196.51 & 288 & 843.98 & 799 & 1360.4 & 791.7679\\
$E$ [GeV] & 6 & 3 & 2 & 2.7 & 6 & 4 & 6 & 3   \\
$\epsilon_{x0}$ [pm-rad] & 42 & 250 & 108 & 158 & 133 & 58 & 34 & 23  \\
$L$ [m] & $\sim$ 6 & \begin{tabular}{@{}c@{}} (LS,SS)= \\ (7.5,6.5) \end{tabular} & $\sim$ 5 & \begin{tabular}{@{}c@{}} (LS,MS,SS)$\sim$ \\ (12,6,4) \end{tabular} & 5.3 & 6.5 & $\sim$ 6 & \begin{tabular}{@{}c@{}} (LS,SS)= \\ (8.4,6.1) \end{tabular}\\
$\alpha_c$ [$\times 10^{-5}$] & 4.04 & 16.3 & 20.3 & 10.5 & 8.51 & 7.86 & 1.56 & 7.76\\
$\sigma_\delta$ [\%] & 0.135 & 0.085 & 0.102 & 0.116 & 0.094 & 0.120 & 0.106 & 0.073\\
$U_0$ [MeV] & 2.870 & 0.475 & 0.245 & 0.688 & 2.56 & 1.010 & 2.890 & 0.196\\
DA [mm/$\sqrt{m}$]\footnote{values on the negative side of dynamic aperture versus $\sqrt{\beta_x}$ at the injection point.} & $\sim$1.8 & $\sim$2.3 & $\sim$0.5 & $\sim$1.7 & $\sim$2.3 & $\sim$3.9 & $\sim$1.3 & $\sim$ 1.1\\
\hline
\end{tabular}
\end{table*}

\section{Conclusions}\label{conclusions}

We presented the complex bend (CB) approach in the lattice design exemplified for the case of the NSLS-II upgrade lattice. 

The benefits of using the CB approach for the high-performance lattice design are demonstrated as follows:
\begin{enumerate}
    \item We achieved a low natural emittance of 23~pm and demonstrated a path of further reducing the emittance by adding more CB poles to the elements in the lattice;
    \item The 8.4-meter-long long straight and 6.1-meter-long short straight provide room for long IDs, and hence help to achieve higher brightness of photon beam; 
    \item A larger drift space in the lattice is available for accelerator equipment, where the fraction of drift space to the total cell length is about 50~\%;
    \item Use of complex bend reduces the number of power supplies, and overall power consumption by the magnet system for the developed NSLS-II upgrade lattice is about 20~\% of that for the present NSLS-II storage ring;
    \item On-going nonlinear dynamics optimization has already indicated appreciable DA and MA. The work is continuing in an effort to secure efficient injection and long beam lifetime.
\end{enumerate}
 
Thus far, the hard-edge model for the complex bend is being used, and more accurate magnet modeling is needed. 

\begin{acknowledgments}
This work used resources of the National Synchrotron
Light Source II, a U.S. Department of Energy (DOE) Office
of Science User Facility operated for the DOE Office of
Science by Brookhaven National Laboratory under
Contract No. DE-SC0012704.
\end{acknowledgments}

\clearpage
\nocite{*}

\bibliography{CB_lattice.bib}

\begin{thebibliography}{33}%
\makeatletter
\providecommand \@ifxundefined [1]{%
 \@ifx{#1\undefined}
}%
\providecommand \@ifnum [1]{%
 \ifnum #1\expandafter \@firstoftwo
 \else \expandafter \@secondoftwo
 \fi
}%
\providecommand \@ifx [1]{%
 \ifx #1\expandafter \@firstoftwo
 \else \expandafter \@secondoftwo
 \fi
}%
\providecommand \natexlab [1]{#1}%
\providecommand \enquote  [1]{``#1''}%
\providecommand \bibnamefont  [1]{#1}%
\providecommand \bibfnamefont [1]{#1}%
\providecommand \citenamefont [1]{#1}%
\providecommand \href@noop [0]{\@secondoftwo}%
\providecommand \href [0]{\begingroup \@sanitize@url \@href}%
\providecommand \@href[1]{\@@startlink{#1}\@@href}%
\providecommand \@@href[1]{\endgroup#1\@@endlink}%
\providecommand \@sanitize@url [0]{\catcode `\\12\catcode `\$12\catcode `\&12\catcode `\#12\catcode `\^12\catcode `\_12\catcode `\%12\relax}%
\providecommand \@@startlink[1]{}%
\providecommand \@@endlink[0]{}%
\providecommand \url  [0]{\begingroup\@sanitize@url \@url }%
\providecommand \@url [1]{\endgroup\@href {#1}{\urlprefix }}%
\providecommand \urlprefix  [0]{URL }%
\providecommand \Eprint [0]{\href }%
\providecommand \doibase [0]{https://doi.org/}%
\providecommand \selectlanguage [0]{\@gobble}%
\providecommand \bibinfo  [0]{\@secondoftwo}%
\providecommand \bibfield  [0]{\@secondoftwo}%
\providecommand \translation [1]{[#1]}%
\providecommand \BibitemOpen [0]{}%
\providecommand \bibitemStop [0]{}%
\providecommand \bibitemNoStop [0]{.\EOS\space}%
\providecommand \EOS [0]{\spacefactor3000\relax}%
\providecommand \BibitemShut  [1]{\csname bibitem#1\endcsname}%
\let\auto@bib@innerbib\@empty
\bibitem [{\citenamefont {Shaftan}\ \emph {et~al.}(2018)\citenamefont {Shaftan}, \citenamefont {Smaluk},\ and\ \citenamefont {Wamg}}]{Timur_2018_techreport}%
  \BibitemOpen
  \bibfield  {author} {\bibinfo {author} {\bibfnamefont {T.}~\bibnamefont {Shaftan}}, \bibinfo {author} {\bibfnamefont {V.}~\bibnamefont {Smaluk}},\ and\ \bibinfo {author} {\bibfnamefont {G.}~\bibnamefont {Wamg}},\ }\href@noop {} {\emph {\bibinfo {title} {Concept of the complex bend}}},\ \bibinfo {type} {Tech. Rep.}\ (\bibinfo  {institution} {Brookhaven National Lab.(BNL), Upton, NY, United States},\ \bibinfo {year} {2018})\BibitemShut {NoStop}%
\bibitem [{\citenamefont {Lee}(2018)}]{Lee_accel_phys_book}%
  \BibitemOpen
  \bibfield  {author} {\bibinfo {author} {\bibfnamefont {S.-Y.}\ \bibnamefont {Lee}},\ }\href@noop {} {\emph {\bibinfo {title} {Accelerator physics}}}\ (\bibinfo  {publisher} {World Scientific Publishing Company},\ \bibinfo {year} {2018})\BibitemShut {NoStop}%
\bibitem [{\citenamefont {Einfeld}\ \emph {et~al.}(1995)\citenamefont {Einfeld}, \citenamefont {Schaper},\ and\ \citenamefont {Plesko}}]{Einfeld_DIFL}%
  \BibitemOpen
  \bibfield  {author} {\bibinfo {author} {\bibfnamefont {D.}~\bibnamefont {Einfeld}}, \bibinfo {author} {\bibfnamefont {J.}~\bibnamefont {Schaper}},\ and\ \bibinfo {author} {\bibfnamefont {M.}~\bibnamefont {Plesko}},\ }\bibfield  {title} {\bibinfo {title} {Design of a diffraction limited light source (difl)},\ }in\ \href {https://doi.org/10.1109/PAC.1995.504602} {\emph {\bibinfo {booktitle} {Proceedings Particle Accelerator Conference}}},\ Vol.~\bibinfo {volume} {1}\ (\bibinfo {year} {1995})\ pp.\ \bibinfo {pages} {177--179 vol.1}\BibitemShut {NoStop}%
\bibitem [{\citenamefont {Leemann}\ \emph {et~al.}(2018)\citenamefont {Leemann}, \citenamefont {Sj{\"o}str{\"o}m},\ and\ \citenamefont {Andersson}}]{Leeman_MAXIV}%
  \BibitemOpen
  \bibfield  {author} {\bibinfo {author} {\bibfnamefont {S.~C.}\ \bibnamefont {Leemann}}, \bibinfo {author} {\bibfnamefont {M.}~\bibnamefont {Sj{\"o}str{\"o}m}},\ and\ \bibinfo {author} {\bibfnamefont {{\AA}.}~\bibnamefont {Andersson}},\ }\bibfield  {title} {\bibinfo {title} {First optics and beam dynamics studies on the {MAX IV} 3 {GeV} storage ring},\ }\href {https://doi.org/https://doi.org/10.1016/j.nima.2017.11.072} {\bibfield  {journal} {\bibinfo  {journal} {Nuclear Instruments and Methods in Physics Research Section A: Accelerators, Spectrometers, Detectors and Associated Equipment}\ }\textbf {\bibinfo {volume} {883}},\ \bibinfo {pages} {33} (\bibinfo {year} {2018})}\BibitemShut {NoStop}%
\bibitem [{\citenamefont {Rodrigues}\ \emph {et~al.}(2018)\citenamefont {Rodrigues} \emph {et~al.}}]{Rodrigues_Sirius}%
  \BibitemOpen
  \bibfield  {author} {\bibinfo {author} {\bibfnamefont {A.}~\bibnamefont {Rodrigues}} \emph {et~al.},\ }\bibfield  {title} {\bibinfo {title} {{S}irius {L}ight {S}ource {S}tatus {R}eport},\ }in\ \href {https://doi.org/doi:10.18429/JACoW-IPAC2018-THXGBD4} {\emph {\bibinfo {booktitle} {Proc. 9th International Particle Accelerator Conference (IPAC'18), Vancouver, BC, Canada, April 29-May 4, 2018}}},\ \bibinfo {series and number} {\bibinfo {series} {International Particle Accelerator Conference}\ No.~\bibinfo {number} {9}}\ (\bibinfo  {publisher} {JACoW Publishing},\ \bibinfo {address} {Geneva, Switzerland},\ \bibinfo {year} {2018})\ pp.\ \bibinfo {pages} {2886--2889},\ \bibinfo {note} {https://doi.org/10.18429/JACoW-IPAC2018-THXGBD4}\BibitemShut {NoStop}%
\bibitem [{\citenamefont {Raimondi}\ \emph {et~al.}(2021)\citenamefont {Raimondi}, \citenamefont {Carmignani}, \citenamefont {Carver}, \citenamefont {Chavanne}, \citenamefont {Farvacque}, \citenamefont {Le~Bec}, \citenamefont {Martin}, \citenamefont {Liuzzo}, \citenamefont {Perron},\ and\ \citenamefont {White}}]{Raimondi_ESRF-EBS}%
  \BibitemOpen
  \bibfield  {author} {\bibinfo {author} {\bibfnamefont {P.}~\bibnamefont {Raimondi}}, \bibinfo {author} {\bibfnamefont {N.}~\bibnamefont {Carmignani}}, \bibinfo {author} {\bibfnamefont {L.~R.}\ \bibnamefont {Carver}}, \bibinfo {author} {\bibfnamefont {J.}~\bibnamefont {Chavanne}}, \bibinfo {author} {\bibfnamefont {L.}~\bibnamefont {Farvacque}}, \bibinfo {author} {\bibfnamefont {G.}~\bibnamefont {Le~Bec}}, \bibinfo {author} {\bibfnamefont {D.}~\bibnamefont {Martin}}, \bibinfo {author} {\bibfnamefont {S.~M.}\ \bibnamefont {Liuzzo}}, \bibinfo {author} {\bibfnamefont {T.}~\bibnamefont {Perron}},\ and\ \bibinfo {author} {\bibfnamefont {S.}~\bibnamefont {White}},\ }\bibfield  {title} {\bibinfo {title} {Commissioning of the hybrid multibend achromat lattice at the european synchrotron radiation facility},\ }\href {https://doi.org/10.1103/PhysRevAccelBeams.24.110701} {\bibfield  {journal} {\bibinfo  {journal} {Phys. Rev. Accel. Beams}\ }\textbf {\bibinfo {volume} {24}},\ \bibinfo {pages} {110701} (\bibinfo {year}
  {2021})}\BibitemShut {NoStop}%
\bibitem [{\citenamefont {Fornek}(2019)}]{Fornek_APS-U}%
  \BibitemOpen
  \bibfield  {author} {\bibinfo {author} {\bibfnamefont {T.~E.}\ \bibnamefont {Fornek}},\ }\bibfield  {title} {\bibinfo {title} {Advanced photon source upgrade project final design report}\ }\href {https://doi.org/10.2172/1543138} {10.2172/1543138} (\bibinfo {year} {2019})\BibitemShut {NoStop}%
\bibitem [{\citenamefont {Steier}\ \emph {et~al.}(2014)\citenamefont {Steier}, \citenamefont {Madur}, \citenamefont {Bailey}, \citenamefont {Berg}, \citenamefont {Biocca}, \citenamefont {Black}, \citenamefont {Casey}, \citenamefont {Colomb}, \citenamefont {Gunion}, \citenamefont {Li}, \citenamefont {Marks}, \citenamefont {Nishimura}, \citenamefont {Pappas}, \citenamefont {Petermann}, \citenamefont {Portmann}, \citenamefont {Prestemon}, \citenamefont {Rawlins}, \citenamefont {Robin}, \citenamefont {Rossi}, \citenamefont {Scarvie}, \citenamefont {Schlueter}, \citenamefont {Sun}, \citenamefont {Tarawneh}, \citenamefont {Wan}, \citenamefont {Williams}, \citenamefont {Yin}, \citenamefont {Zhou}, \citenamefont {Jin}, \citenamefont {Zhang}, \citenamefont {Chen}, \citenamefont {Wen},\ and\ \citenamefont {Wu}}]{Steier_ALS-U}%
  \BibitemOpen
  \bibfield  {author} {\bibinfo {author} {\bibfnamefont {C.}~\bibnamefont {Steier}}, \bibinfo {author} {\bibfnamefont {A.}~\bibnamefont {Madur}}, \bibinfo {author} {\bibfnamefont {B.}~\bibnamefont {Bailey}}, \bibinfo {author} {\bibfnamefont {K.}~\bibnamefont {Berg}}, \bibinfo {author} {\bibfnamefont {A.}~\bibnamefont {Biocca}}, \bibinfo {author} {\bibfnamefont {A.}~\bibnamefont {Black}}, \bibinfo {author} {\bibfnamefont {P.}~\bibnamefont {Casey}}, \bibinfo {author} {\bibfnamefont {D.}~\bibnamefont {Colomb}}, \bibinfo {author} {\bibfnamefont {B.}~\bibnamefont {Gunion}}, \bibinfo {author} {\bibfnamefont {N.}~\bibnamefont {Li}}, \bibinfo {author} {\bibfnamefont {S.}~\bibnamefont {Marks}}, \bibinfo {author} {\bibfnamefont {H.}~\bibnamefont {Nishimura}}, \bibinfo {author} {\bibfnamefont {C.}~\bibnamefont {Pappas}}, \bibinfo {author} {\bibfnamefont {K.}~\bibnamefont {Petermann}}, \bibinfo {author} {\bibfnamefont {G.}~\bibnamefont {Portmann}}, \bibinfo {author} {\bibfnamefont {S.}~\bibnamefont {Prestemon}}, \bibinfo
  {author} {\bibfnamefont {A.}~\bibnamefont {Rawlins}}, \bibinfo {author} {\bibfnamefont {D.}~\bibnamefont {Robin}}, \bibinfo {author} {\bibfnamefont {S.}~\bibnamefont {Rossi}}, \bibinfo {author} {\bibfnamefont {T.}~\bibnamefont {Scarvie}}, \bibinfo {author} {\bibfnamefont {R.}~\bibnamefont {Schlueter}}, \bibinfo {author} {\bibfnamefont {C.}~\bibnamefont {Sun}}, \bibinfo {author} {\bibfnamefont {H.}~\bibnamefont {Tarawneh}}, \bibinfo {author} {\bibfnamefont {W.}~\bibnamefont {Wan}}, \bibinfo {author} {\bibfnamefont {E.}~\bibnamefont {Williams}}, \bibinfo {author} {\bibfnamefont {L.}~\bibnamefont {Yin}}, \bibinfo {author} {\bibfnamefont {Q.}~\bibnamefont {Zhou}}, \bibinfo {author} {\bibfnamefont {J.}~\bibnamefont {Jin}}, \bibinfo {author} {\bibfnamefont {J.}~\bibnamefont {Zhang}}, \bibinfo {author} {\bibfnamefont {C.}~\bibnamefont {Chen}}, \bibinfo {author} {\bibfnamefont {Y.}~\bibnamefont {Wen}},\ and\ \bibinfo {author} {\bibfnamefont {J.}~\bibnamefont {Wu}},\ }\bibfield  {title} {\bibinfo {title} {Completion
  of the brightness upgrade of the {ALS}},\ }\href {https://doi.org/10.1088/1742-6596/493/1/012030} {\bibfield  {journal} {\bibinfo  {journal} {Journal of Physics: Conference Series}\ }\textbf {\bibinfo {volume} {493}},\ \bibinfo {pages} {012030} (\bibinfo {year} {2014})}\BibitemShut {NoStop}%
\bibitem [{\citenamefont {Hellert}\ \emph {et~al.}(2022)\citenamefont {Hellert}, \citenamefont {Steier},\ and\ \citenamefont {Venturini}}]{Hellert_ALS-U}%
  \BibitemOpen
  \bibfield  {author} {\bibinfo {author} {\bibfnamefont {T.}~\bibnamefont {Hellert}}, \bibinfo {author} {\bibfnamefont {C.}~\bibnamefont {Steier}},\ and\ \bibinfo {author} {\bibfnamefont {M.}~\bibnamefont {Venturini}},\ }\bibfield  {title} {\bibinfo {title} {Lattice correction and commissioning simulation of the advanced light source upgrade storage ring},\ }\href {https://doi.org/10.1103/PhysRevAccelBeams.25.110701} {\bibfield  {journal} {\bibinfo  {journal} {Phys. Rev. Accel. Beams}\ }\textbf {\bibinfo {volume} {25}},\ \bibinfo {pages} {110701} (\bibinfo {year} {2022})}\BibitemShut {NoStop}%
\bibitem [{\citenamefont {Jiao}\ \emph {et~al.}(2018)\citenamefont {Jiao}, \citenamefont {Xu}, \citenamefont {Cui}, \citenamefont {Duan}, \citenamefont {Guo}, \citenamefont {He}, \citenamefont {Ji}, \citenamefont {Li}, \citenamefont {Li}, \citenamefont {Meng}, \citenamefont {Peng}, \citenamefont {Tian}, \citenamefont {Wang}, \citenamefont {Wang}, \citenamefont {Wei}, \citenamefont {Xu}, \citenamefont {Yan}, \citenamefont {Yu}, \citenamefont {Zhao},\ and\ \citenamefont {Qin}}]{Jiao_HEPS}%
  \BibitemOpen
  \bibfield  {author} {\bibinfo {author} {\bibfnamefont {Y.}~\bibnamefont {Jiao}}, \bibinfo {author} {\bibfnamefont {G.}~\bibnamefont {Xu}}, \bibinfo {author} {\bibfnamefont {X.-H.}\ \bibnamefont {Cui}}, \bibinfo {author} {\bibfnamefont {Z.}~\bibnamefont {Duan}}, \bibinfo {author} {\bibfnamefont {Y.-Y.}\ \bibnamefont {Guo}}, \bibinfo {author} {\bibfnamefont {P.}~\bibnamefont {He}}, \bibinfo {author} {\bibfnamefont {D.-H.}\ \bibnamefont {Ji}}, \bibinfo {author} {\bibfnamefont {J.-Y.}\ \bibnamefont {Li}}, \bibinfo {author} {\bibfnamefont {X.-Y.}\ \bibnamefont {Li}}, \bibinfo {author} {\bibfnamefont {C.}~\bibnamefont {Meng}}, \bibinfo {author} {\bibfnamefont {Y.-M.}\ \bibnamefont {Peng}}, \bibinfo {author} {\bibfnamefont {S.-K.}\ \bibnamefont {Tian}}, \bibinfo {author} {\bibfnamefont {J.-Q.}\ \bibnamefont {Wang}}, \bibinfo {author} {\bibfnamefont {N.}~\bibnamefont {Wang}}, \bibinfo {author} {\bibfnamefont {Y.-Y.}\ \bibnamefont {Wei}}, \bibinfo {author} {\bibfnamefont {H.-S.}\ \bibnamefont {Xu}}, \bibinfo
  {author} {\bibfnamefont {F.}~\bibnamefont {Yan}}, \bibinfo {author} {\bibfnamefont {C.-H.}\ \bibnamefont {Yu}}, \bibinfo {author} {\bibfnamefont {Y.-L.}\ \bibnamefont {Zhao}},\ and\ \bibinfo {author} {\bibfnamefont {Q.}~\bibnamefont {Qin}},\ }\bibfield  {title} {\bibinfo {title} {{The HEPS project}},\ }\href {https://doi.org/10.1107/S1600577518012110} {\bibfield  {journal} {\bibinfo  {journal} {Journal of Synchrotron Radiation}\ }\textbf {\bibinfo {volume} {25}},\ \bibinfo {pages} {1611} (\bibinfo {year} {2018})}\BibitemShut {NoStop}%
\bibitem [{\citenamefont {Raimondi}\ and\ \citenamefont {Liuzzo}(2023)}]{Raimondi_DLS}%
  \BibitemOpen
  \bibfield  {author} {\bibinfo {author} {\bibfnamefont {P.}~\bibnamefont {Raimondi}}\ and\ \bibinfo {author} {\bibfnamefont {S.~M.}\ \bibnamefont {Liuzzo}},\ }\bibfield  {title} {\bibinfo {title} {Toward a diffraction limited light source},\ }\href {https://doi.org/10.1103/PhysRevAccelBeams.26.021601} {\bibfield  {journal} {\bibinfo  {journal} {Phys. Rev. Accel. Beams}\ }\textbf {\bibinfo {volume} {26}},\ \bibinfo {pages} {021601} (\bibinfo {year} {2023})}\BibitemShut {NoStop}%
\bibitem [{\citenamefont {Streun}\ \emph {et~al.}(2023)\citenamefont {Streun}, \citenamefont {Aiba}, \citenamefont {B\"oge}, \citenamefont {Calzolaio}, \citenamefont {Ehrlichman}, \citenamefont {Negrazus}, \citenamefont {Riemann},\ and\ \citenamefont {Vrankovic}}]{Streun_SLSII}%
  \BibitemOpen
  \bibfield  {author} {\bibinfo {author} {\bibfnamefont {A.}~\bibnamefont {Streun}}, \bibinfo {author} {\bibfnamefont {M.}~\bibnamefont {Aiba}}, \bibinfo {author} {\bibfnamefont {M.}~\bibnamefont {B\"oge}}, \bibinfo {author} {\bibfnamefont {C.}~\bibnamefont {Calzolaio}}, \bibinfo {author} {\bibfnamefont {M.}~\bibnamefont {Ehrlichman}}, \bibinfo {author} {\bibfnamefont {M.}~\bibnamefont {Negrazus}}, \bibinfo {author} {\bibfnamefont {B.}~\bibnamefont {Riemann}},\ and\ \bibinfo {author} {\bibfnamefont {V.}~\bibnamefont {Vrankovic}},\ }\bibfield  {title} {\bibinfo {title} {Swiss light source upgrade lattice design},\ }\href {https://doi.org/10.1103/PhysRevAccelBeams.26.091601} {\bibfield  {journal} {\bibinfo  {journal} {Phys. Rev. Accel. Beams}\ }\textbf {\bibinfo {volume} {26}},\ \bibinfo {pages} {091601} (\bibinfo {year} {2023})}\BibitemShut {NoStop}%
\bibitem [{\citenamefont {Bartolini}\ \emph {et~al.}(2018)\citenamefont {Bartolini}, \citenamefont {Abraham}, \citenamefont {Apollonio}, \citenamefont {Bailey}, \citenamefont {Cox}, \citenamefont {Day}, \citenamefont {Fielder}, \citenamefont {Hammond}, \citenamefont {Heron}, \citenamefont {Holdsworth}, \citenamefont {Kay}, \citenamefont {Martin}, \citenamefont {Mhaskar}, \citenamefont {Miller}, \citenamefont {Pulampong}, \citenamefont {Rehm}, \citenamefont {Rial}, \citenamefont {Rose}, \citenamefont {Shahveh}, \citenamefont {Singh}, \citenamefont {Thomson},\ and\ \citenamefont {Walker}}]{Bartolini_Diamond-U}%
  \BibitemOpen
  \bibfield  {author} {\bibinfo {author} {\bibfnamefont {R.}~\bibnamefont {Bartolini}}, \bibinfo {author} {\bibfnamefont {C.}~\bibnamefont {Abraham}}, \bibinfo {author} {\bibfnamefont {M.}~\bibnamefont {Apollonio}}, \bibinfo {author} {\bibfnamefont {C.~P.}\ \bibnamefont {Bailey}}, \bibinfo {author} {\bibfnamefont {M.~P.}\ \bibnamefont {Cox}}, \bibinfo {author} {\bibfnamefont {A.}~\bibnamefont {Day}}, \bibinfo {author} {\bibfnamefont {R.~T.}\ \bibnamefont {Fielder}}, \bibinfo {author} {\bibfnamefont {N.~P.}\ \bibnamefont {Hammond}}, \bibinfo {author} {\bibfnamefont {M.~T.}\ \bibnamefont {Heron}}, \bibinfo {author} {\bibfnamefont {R.}~\bibnamefont {Holdsworth}}, \bibinfo {author} {\bibfnamefont {J.}~\bibnamefont {Kay}}, \bibinfo {author} {\bibfnamefont {I.~P.~S.}\ \bibnamefont {Martin}}, \bibinfo {author} {\bibfnamefont {S.}~\bibnamefont {Mhaskar}}, \bibinfo {author} {\bibfnamefont {A.}~\bibnamefont {Miller}}, \bibinfo {author} {\bibfnamefont {T.}~\bibnamefont {Pulampong}}, \bibinfo {author} {\bibfnamefont
  {G.}~\bibnamefont {Rehm}}, \bibinfo {author} {\bibfnamefont {E.~C.~M.}\ \bibnamefont {Rial}}, \bibinfo {author} {\bibfnamefont {A.}~\bibnamefont {Rose}}, \bibinfo {author} {\bibfnamefont {A.}~\bibnamefont {Shahveh}}, \bibinfo {author} {\bibfnamefont {B.}~\bibnamefont {Singh}}, \bibinfo {author} {\bibfnamefont {A.}~\bibnamefont {Thomson}},\ and\ \bibinfo {author} {\bibfnamefont {R.~P.}\ \bibnamefont {Walker}},\ }\bibfield  {title} {\bibinfo {title} {Double-double bend achromat cell upgrade at the diamond light source: From design to commissioning},\ }\href {https://doi.org/10.1103/PhysRevAccelBeams.21.050701} {\bibfield  {journal} {\bibinfo  {journal} {Phys. Rev. Accel. Beams}\ }\textbf {\bibinfo {volume} {21}},\ \bibinfo {pages} {050701} (\bibinfo {year} {2018})}\BibitemShut {NoStop}%
\bibitem [{\citenamefont {Karantzoulis}(2018)}]{ElettraII}%
  \BibitemOpen
  \bibfield  {author} {\bibinfo {author} {\bibfnamefont {E.}~\bibnamefont {Karantzoulis}},\ }\bibfield  {title} {\bibinfo {title} {Elettra 2.0 — the diffraction limited successor of elettra},\ }\href {https://doi.org/https://doi.org/10.1016/j.nima.2017.09.057} {\bibfield  {journal} {\bibinfo  {journal} {Nuclear Instruments and Methods in Physics Research Section A: Accelerators, Spectrometers, Detectors and Associated Equipment}\ }\textbf {\bibinfo {volume} {880}},\ \bibinfo {pages} {158} (\bibinfo {year} {2018})}\BibitemShut {NoStop}%
\bibitem [{\citenamefont {Tavares}\ \emph {et~al.}(2018)\citenamefont {Tavares}, \citenamefont {Bengtsson},\ and\ \citenamefont {Åke Andersson}}]{Tavares_MAXIV_future}%
  \BibitemOpen
  \bibfield  {author} {\bibinfo {author} {\bibfnamefont {P.~F.}\ \bibnamefont {Tavares}}, \bibinfo {author} {\bibfnamefont {J.}~\bibnamefont {Bengtsson}},\ and\ \bibinfo {author} {\bibnamefont {Åke Andersson}},\ }\bibfield  {title} {\bibinfo {title} {Future development plans for the max iv light source: Pushing further towards higher brightness and coherence},\ }\href {https://doi.org/https://doi.org/10.1016/j.elspec.2017.09.010} {\bibfield  {journal} {\bibinfo  {journal} {Journal of Electron Spectroscopy and Related Phenomena}\ }\textbf {\bibinfo {volume} {224}},\ \bibinfo {pages} {8} (\bibinfo {year} {2018})}\BibitemShut {NoStop}%
\bibitem [{\citenamefont {Sharma}(2023)}]{Sushil_retreat_slides}%
  \BibitemOpen
  \bibfield  {author} {\bibinfo {author} {\bibfnamefont {S.}~\bibnamefont {Sharma}},\ }\href@noop {} {\bibinfo {title} {Elements of engineering design of the upgrade lattice}},\ \bibinfo {howpublished} {Retreat on Plans for NSLS-II Upgrade} (\bibinfo {year} {2023})\BibitemShut {NoStop}%
\bibitem [{\citenamefont {Brooks}(2021)}]{Brooks_BNL_PMQ}%
  \BibitemOpen
  \bibfield  {author} {\bibinfo {author} {\bibfnamefont {S.}~\bibnamefont {Brooks}},\ }\bibfield  {title} {\bibinfo {title} {{Modified Halbach Magnets for Emerging Accelerator Applications}},\ }in\ \href {https://doi.org/10.18429/JACoW-IPAC2021-TUXC07} {\emph {\bibinfo {booktitle} {Proc. IPAC'21}}},\ \bibinfo {series and number} {\bibinfo {series} {International Particle Accelerator Conference}\ No.~\bibinfo {number} {12}}\ (\bibinfo  {publisher} {JACoW Publishing, Geneva, Switzerland},\ \bibinfo {year} {2021})\ pp.\ \bibinfo {pages} {1315--1318},\ \bibinfo {note} {https://doi.org/10.18429/JACoW-IPAC2021-TUXC07}\BibitemShut {NoStop}%
\bibitem [{\citenamefont {N'gotta}\ \emph {et~al.}(2016)\citenamefont {N'gotta}, \citenamefont {Le~Bec},\ and\ \citenamefont {Chavanne}}]{N'gotta_ESRF_PMQ}%
  \BibitemOpen
  \bibfield  {author} {\bibinfo {author} {\bibfnamefont {P.}~\bibnamefont {N'gotta}}, \bibinfo {author} {\bibfnamefont {G.}~\bibnamefont {Le~Bec}},\ and\ \bibinfo {author} {\bibfnamefont {J.}~\bibnamefont {Chavanne}},\ }\bibfield  {title} {\bibinfo {title} {Hybrid high gradient permanent magnet quadrupole},\ }\href {https://doi.org/10.1103/PhysRevAccelBeams.19.122401} {\bibfield  {journal} {\bibinfo  {journal} {Phys. Rev. Accel. Beams}\ }\textbf {\bibinfo {volume} {19}},\ \bibinfo {pages} {122401} (\bibinfo {year} {2016})}\BibitemShut {NoStop}%
\bibitem [{\citenamefont {Ghaith}\ \emph {et~al.}(2019)\citenamefont {Ghaith}, \citenamefont {Oumbarek}, \citenamefont {Kitégi}, \citenamefont {Valléau}, \citenamefont {Marteau},\ and\ \citenamefont {Couprie}}]{Ghaith_SOLEIL_PMQ}%
  \BibitemOpen
  \bibfield  {author} {\bibinfo {author} {\bibfnamefont {A.}~\bibnamefont {Ghaith}}, \bibinfo {author} {\bibfnamefont {D.}~\bibnamefont {Oumbarek}}, \bibinfo {author} {\bibfnamefont {C.}~\bibnamefont {Kitégi}}, \bibinfo {author} {\bibfnamefont {M.}~\bibnamefont {Valléau}}, \bibinfo {author} {\bibfnamefont {F.}~\bibnamefont {Marteau}},\ and\ \bibinfo {author} {\bibfnamefont {M.-E.}\ \bibnamefont {Couprie}},\ }\bibfield  {title} {\bibinfo {title} {Permanent magnet-based quadrupoles for plasma acceleration sources},\ }\bibfield  {journal} {\bibinfo  {journal} {Instruments}\ }\textbf {\bibinfo {volume} {3}},\ \href {https://doi.org/10.3390/instruments3020027} {10.3390/instruments3020027} (\bibinfo {year} {2019})\BibitemShut {NoStop}%
\bibitem [{\citenamefont {Shaftan}(2022)}]{Timur_Methodology_TCBA}%
  \BibitemOpen
  \bibfield  {author} {\bibinfo {author} {\bibfnamefont {T.}~\bibnamefont {Shaftan}},\ }\href@noop {} {\emph {\bibinfo {title} {Methodology for designing Complex Bend lattice}}},\ \bibinfo {type} {Tech. Rep.}\ (\bibinfo  {institution} {Brookhaven National Lab.(BNL), Upton, NY, United States},\ \bibinfo {year} {2022})\BibitemShut {NoStop}%
\bibitem [{\citenamefont {Wang}\ \emph {et~al.}(2018)\citenamefont {Wang}, \citenamefont {Shaftan}, \citenamefont {Smaluk}, \citenamefont {Mezentsev}, \citenamefont {Sharma}, \citenamefont {Chubar}, \citenamefont {Hidaka},\ and\ \citenamefont {Spataro}}]{Guimei_CBI}%
  \BibitemOpen
  \bibfield  {author} {\bibinfo {author} {\bibfnamefont {G.}~\bibnamefont {Wang}}, \bibinfo {author} {\bibfnamefont {T.}~\bibnamefont {Shaftan}}, \bibinfo {author} {\bibfnamefont {V.}~\bibnamefont {Smaluk}}, \bibinfo {author} {\bibfnamefont {N.~A.}\ \bibnamefont {Mezentsev}}, \bibinfo {author} {\bibfnamefont {S.}~\bibnamefont {Sharma}}, \bibinfo {author} {\bibfnamefont {O.}~\bibnamefont {Chubar}}, \bibinfo {author} {\bibfnamefont {Y.}~\bibnamefont {Hidaka}},\ and\ \bibinfo {author} {\bibfnamefont {C.}~\bibnamefont {Spataro}},\ }\bibfield  {title} {\bibinfo {title} {Complex bend: Strong-focusing magnet for low-emittance synchrotrons},\ }\href {https://doi.org/10.1103/PhysRevAccelBeams.21.100703} {\bibfield  {journal} {\bibinfo  {journal} {Phys. Rev. Accel. Beams}\ }\textbf {\bibinfo {volume} {21}},\ \bibinfo {pages} {100703} (\bibinfo {year} {2018})}\BibitemShut {NoStop}%
\bibitem [{\citenamefont {Wang}\ \emph {et~al.}(2019)\citenamefont {Wang}, \citenamefont {Shaftan}, \citenamefont {Smaluk}, \citenamefont {Hidaka}, \citenamefont {Chubar}, \citenamefont {Tanabe}, \citenamefont {Choi}, \citenamefont {Sharma}, \citenamefont {Spataro},\ and\ \citenamefont {Mesentsev}}]{Guimei_CBII}%
  \BibitemOpen
  \bibfield  {author} {\bibinfo {author} {\bibfnamefont {G.}~\bibnamefont {Wang}}, \bibinfo {author} {\bibfnamefont {T.}~\bibnamefont {Shaftan}}, \bibinfo {author} {\bibfnamefont {V.}~\bibnamefont {Smaluk}}, \bibinfo {author} {\bibfnamefont {Y.}~\bibnamefont {Hidaka}}, \bibinfo {author} {\bibfnamefont {O.}~\bibnamefont {Chubar}}, \bibinfo {author} {\bibfnamefont {T.}~\bibnamefont {Tanabe}}, \bibinfo {author} {\bibfnamefont {J.}~\bibnamefont {Choi}}, \bibinfo {author} {\bibfnamefont {S.}~\bibnamefont {Sharma}}, \bibinfo {author} {\bibfnamefont {C.}~\bibnamefont {Spataro}},\ and\ \bibinfo {author} {\bibfnamefont {N.~A.}\ \bibnamefont {Mesentsev}},\ }\bibfield  {title} {\bibinfo {title} {Complex bend. ii. a new optics solution},\ }\href {https://doi.org/10.1103/PhysRevAccelBeams.22.110703} {\bibfield  {journal} {\bibinfo  {journal} {Phys. Rev. Accel. Beams}\ }\textbf {\bibinfo {volume} {22}},\ \bibinfo {pages} {110703} (\bibinfo {year} {2019})}\BibitemShut {NoStop}%
\bibitem [{\citenamefont {Iselin}\ and\ \citenamefont {Grote}(1995)}]{Iselin_MAD8}%
  \BibitemOpen
  \bibfield  {author} {\bibinfo {author} {\bibfnamefont {F.~C.}\ \bibnamefont {Iselin}}\ and\ \bibinfo {author} {\bibfnamefont {H.}~\bibnamefont {Grote}},\ }\href@noop {} {\emph {\bibinfo {title} {The MAD Program (Methodical Accelerator Design) Version 8.13/8 User’s Reference Manual}}},\ \bibinfo {type} {Tech. Rep.}\ (\bibinfo  {institution} {CERN},\ \bibinfo {year} {1995})\BibitemShut {NoStop}%
\bibitem [{\citenamefont {Inc.}(2023)}]{matlab}%
  \BibitemOpen
  \bibfield  {author} {\bibinfo {author} {\bibfnamefont {T.~M.}\ \bibnamefont {Inc.}},\ }\href {https://www.mathworks.com} {\bibinfo {title} {Matlab version: 9.14.0.2306882 (r2023a)}} (\bibinfo {year} {2023})\BibitemShut {NoStop}%
\bibitem [{\citenamefont {{Deb}}\ \emph {et~al.}(2002)\citenamefont {{Deb}}, \citenamefont {{Pratap}}, \citenamefont {{Agarwal}},\ and\ \citenamefont {{Meyarivan}}}]{Deb_NSGAII}%
  \BibitemOpen
  \bibfield  {author} {\bibinfo {author} {\bibfnamefont {K.}~\bibnamefont {{Deb}}}, \bibinfo {author} {\bibfnamefont {A.}~\bibnamefont {{Pratap}}}, \bibinfo {author} {\bibfnamefont {S.}~\bibnamefont {{Agarwal}}},\ and\ \bibinfo {author} {\bibfnamefont {T.}~\bibnamefont {{Meyarivan}}},\ }\bibfield  {title} {\bibinfo {title} {A fast and elitist multiobjective genetic algorithm: {NSGA-II}},\ }\href {https://doi.org/10.1109/4235.996017} {\bibfield  {journal} {\bibinfo  {journal} {IEEE Transactions on Evolutionary Computation}\ }\textbf {\bibinfo {volume} {6}},\ \bibinfo {pages} {182} (\bibinfo {year} {2002})}\BibitemShut {NoStop}%
\bibitem [{\citenamefont {{Aravind Seshadri (2020)}}()}]{matlab_NSGA-II}%
  \BibitemOpen
  \bibfield  {author} {\bibinfo {author} {\bibnamefont {{Aravind Seshadri (2020)}}},\ }\href@noop {} {\bibinfo {title} {{NSGA-II}: A multi-objective optimization algorithm}},\ \bibinfo {note} {{(https://www.mathworks.com/matlabcentral/fileexchange/10429-nsga-ii-a-multi-objective-optimization-algorithm), MATLAB Central File Exchange. Retrieved January 6, 2020.}}\BibitemShut {Stop}%
\bibitem [{\citenamefont {McKay}\ \emph {et~al.}(1979)\citenamefont {McKay}, \citenamefont {Beckman},\ and\ \citenamefont {Conover}}]{LHS}%
  \BibitemOpen
  \bibfield  {author} {\bibinfo {author} {\bibfnamefont {M.~D.}\ \bibnamefont {McKay}}, \bibinfo {author} {\bibfnamefont {R.~J.}\ \bibnamefont {Beckman}},\ and\ \bibinfo {author} {\bibfnamefont {W.~J.}\ \bibnamefont {Conover}},\ }\bibfield  {title} {\bibinfo {title} {A comparison of three methods for selecting values of input variables in the analysis of output from a computer code},\ }\href {http://www.jstor.org/stable/1268522} {\bibfield  {journal} {\bibinfo  {journal} {Technometrics}\ }\textbf {\bibinfo {volume} {21}},\ \bibinfo {pages} {239} (\bibinfo {year} {1979})}\BibitemShut {NoStop}%
\bibitem [{\citenamefont {Helm}\ \emph {et~al.}(1973)\citenamefont {Helm}, \citenamefont {Lee}, \citenamefont {Morton},\ and\ \citenamefont {Sands}}]{Helm_radiation_integrals}%
  \BibitemOpen
  \bibfield  {author} {\bibinfo {author} {\bibfnamefont {R.~H.}\ \bibnamefont {Helm}}, \bibinfo {author} {\bibfnamefont {M.~J.}\ \bibnamefont {Lee}}, \bibinfo {author} {\bibfnamefont {P.}~\bibnamefont {Morton}},\ and\ \bibinfo {author} {\bibfnamefont {M.}~\bibnamefont {Sands}},\ }\bibfield  {title} {\bibinfo {title} {Evaluation of synchrotron radiation integrals},\ }\href@noop {} {\bibfield  {journal} {\bibinfo  {journal} {IEEE Trans. Nucl. Sci}\ }\textbf {\bibinfo {volume} {20}},\ \bibinfo {pages} {43} (\bibinfo {year} {1973})}\BibitemShut {NoStop}%
\bibitem [{\citenamefont {Plassard}\ \emph {et~al.}(2021)\citenamefont {Plassard}, \citenamefont {Wang}, \citenamefont {Shaftan}, \citenamefont {Smaluk}, \citenamefont {Li},\ and\ \citenamefont {Hidaka}}]{Fabien_octupole_correction}%
  \BibitemOpen
  \bibfield  {author} {\bibinfo {author} {\bibfnamefont {F.}~\bibnamefont {Plassard}}, \bibinfo {author} {\bibfnamefont {G.}~\bibnamefont {Wang}}, \bibinfo {author} {\bibfnamefont {T.}~\bibnamefont {Shaftan}}, \bibinfo {author} {\bibfnamefont {V.}~\bibnamefont {Smaluk}}, \bibinfo {author} {\bibfnamefont {Y.}~\bibnamefont {Li}},\ and\ \bibinfo {author} {\bibfnamefont {Y.}~\bibnamefont {Hidaka}},\ }\bibfield  {title} {\bibinfo {title} {Simultaneous correction of high order geometrical driving terms with octupoles in synchrotron light sources},\ }\href {https://doi.org/10.1103/PhysRevAccelBeams.24.114801} {\bibfield  {journal} {\bibinfo  {journal} {Phys. Rev. Accel. Beams}\ }\textbf {\bibinfo {volume} {24}},\ \bibinfo {pages} {114801} (\bibinfo {year} {2021})}\BibitemShut {NoStop}%
\bibitem [{\citenamefont {Song}\ and\ \citenamefont {Li}(2023)}]{Minghao_retreat_slides}%
  \BibitemOpen
  \bibfield  {author} {\bibinfo {author} {\bibfnamefont {M.}~\bibnamefont {Song}}\ and\ \bibinfo {author} {\bibfnamefont {Y.}~\bibnamefont {Li}},\ }\href@noop {} {\bibinfo {title} {Low emittance ring lattice design}},\ \bibinfo {howpublished} {Retreat on Plans for NSLS-II Upgrade} (\bibinfo {year} {2023})\BibitemShut {NoStop}%
\bibitem [{\citenamefont {Jang}\ \emph {et~al.}(2022)\citenamefont {Jang}, \citenamefont {Shin}, \citenamefont {Yoon}, \citenamefont {Ko}, \citenamefont {Yoon}, \citenamefont {Lee},\ and\ \citenamefont {Oh}}]{Jang_Korea-4GSR}%
  \BibitemOpen
  \bibfield  {author} {\bibinfo {author} {\bibfnamefont {G.}~\bibnamefont {Jang}}, \bibinfo {author} {\bibfnamefont {S.}~\bibnamefont {Shin}}, \bibinfo {author} {\bibfnamefont {M.}~\bibnamefont {Yoon}}, \bibinfo {author} {\bibfnamefont {J.}~\bibnamefont {Ko}}, \bibinfo {author} {\bibfnamefont {Y.~D.}\ \bibnamefont {Yoon}}, \bibinfo {author} {\bibfnamefont {J.}~\bibnamefont {Lee}},\ and\ \bibinfo {author} {\bibfnamefont {B.-H.}\ \bibnamefont {Oh}},\ }\bibfield  {title} {\bibinfo {title} {Low emittance lattice design for korea-4gsr},\ }\href {https://doi.org/https://doi.org/10.1016/j.nima.2022.166779} {\bibfield  {journal} {\bibinfo  {journal} {Nuclear Instruments and Methods in Physics Research Section A: Accelerators, Spectrometers, Detectors and Associated Equipment}\ }\textbf {\bibinfo {volume} {1034}},\ \bibinfo {pages} {166779} (\bibinfo {year} {2022})}\BibitemShut {NoStop}%
\bibitem [{\citenamefont {Borland}(2000)}]{Borland_elegant}%
  \BibitemOpen
  \bibfield  {author} {\bibinfo {author} {\bibfnamefont {M.}~\bibnamefont {Borland}},\ }\bibfield  {title} {\bibinfo {title} {{elegant: A Flexible SDDS-Compliant Code for Accelerator Simulation}},\ }in\ \href {https://doi.org/10.2172/761286} {\emph {\bibinfo {booktitle} {{6th International Computational Accelerator Physics Conference (ICAP 2000)}}}}\ (\bibinfo {year} {2000})\BibitemShut {NoStop}%
\bibitem [{\citenamefont {Borland}(2010)}]{Borland_swapout}%
  \BibitemOpen
  \bibfield  {author} {\bibinfo {author} {\bibfnamefont {M.}~\bibnamefont {Borland}},\ }\bibfield  {title} {\bibinfo {title} {{Concepts and performance for a next‐generation storage ring hard x‐ray source}},\ }\href {https://doi.org/10.1063/1.3463364} {\bibfield  {journal} {\bibinfo  {journal} {AIP Conference Proceedings}\ }\textbf {\bibinfo {volume} {1234}},\ \bibinfo {pages} {911} (\bibinfo {year} {2010})},\ \Eprint {https://arxiv.org/abs/https://pubs.aip.org/aip/acp/article-pdf/1234/1/911/11959744/911\_1\_online.pdf} {https://pubs.aip.org/aip/acp/article-pdf/1234/1/911/11959744/911\_1\_online.pdf} \BibitemShut {NoStop}%
\end{thebibliography}%

\end{document}